\documentclass{article}
\usepackage[bottom=0.9in]{geometry}
\usepackage{listings}             
\usepackage[affil-it]{authblk}
\usepackage{graphicx}
\usepackage{lscape}
\usepackage[dvipsnames]{xcolor}
\usepackage{caption}
\usepackage{amsmath}
\usepackage{dblfloatfix}
\usepackage{subfigure}
\usepackage{float}
\usepackage{listings}
\usepackage{epstopdf}
\usepackage{url}
\usepackage{cite}
\author{Andrejs Tucs$^1$, Koji Tsuda$^{1,2,3}$ and Adnan Sljoka$^{2,*}$}
\date{%
    $^1$Graduate School of Frontier Sciences, The University of Tokyo, Kashiwa, Chiba 277-8561, Japan\\%
    $^2$RIKEN Center for Advanced Intelligence Project, Tokyo 103-0027\\
    $^3$Research and Services Division of Materials Data and Integrated System, National Institute for Materials Science, Tsukuba, Ibaraki 305-0047, Japan\\
    $^*$adnan.sljoka@riken.jp\\
[2ex]%
    \today
}

\title{Probing conformational dynamics of antibodies with geometric simulations}

\providecommand{\keywords}[1]{\textbf{Key words } #1}

\begin{document}

\maketitle
\begin{abstract}
This chapter describes the application of constrained geometric simulations for prediction of antibody structural dynamics. We utilize constrained geometric simulations method FRODAN, which is a low computational complexity alternative to Molecular Dynamics (MD) simulations that can rapidly explore flexible motions in protein structures. FRODAN is highly suited for conformational dynamics analysis of large proteins, complexes, intrinsically disordered proteins and dynamics that occurs on longer biologically relevant time scales which are normally inaccessible to classical MD simulations. This approach predicts protein dynamics at an all-atom scale while retaining realistic covalent bonding, maintaining dihedral angles in energetically good conformations while avoiding steric clashes in addition to performing other geometric and stereochemical criteria checks. In this chapter, we apply FRODAN to showcase its applicability for probing functionally relevant dynamics of IgG2a, including large amplitude domain-domain motions and motions of complementarity determining region (CDR) loops. As was suggested in previous experimental studies, our simulations show that antibodies can explore a large range of conformational space.
\end{abstract}


\section{Introduction}

In order to understand how proteins carry out their function, we are required to elucidate their structural flexibility and dynamics at atomistic level. Accurate measurements of protein flexibility and dynamics can help us interpret the relationship between structure and function, with implications in drug discovery. While protein structure determination methods such as X-ray crystallography, Nuclear Magnetic Resonance and recent renaissance in cryo-electron microscopy have provided many high-resolution structural snapshots of proteins, they rarely disclose information about intermediate states and functionally relevant conformational flexibility and dynamics. Shifts in conformational dynamics can play a critical role in complexation such as antibody-antigen, protein-protein such as GPCR-G-protein interactions, in understanding complex allosteric effects and the subtle dynamic alterations that modulate binding affinity and specificity.

Determining flexible and rigid regions in proteins and understanding their motions is a complex task as conformational fluctuations often involve a large number of internal degrees of freedom. The fluctuations can be rapid, transient and result in structures that are spectroscopically indistinguishable from the ground-state. To explore motions of highly dynamic and flexible proteins, we need to go beyond structural snapshot representations, typically obtained from structures deposited in Protein Data Bank. While protein dynamics and fast conformational fluctuations can be probed with molecular dynamics (MD) simulations, MD simulations particularly on large protein structures and structures with high disorder are still largely impractical as it takes a prohibitive amount of computational power to investigate functionally relevant motions occurring on the longer micro to milli-second time-scales. The computational time needed to reach motions on these longer timescales, even with massively parallelized MD runs via costly special-purpose commodity computer clusters, is still beyond practical wide-range application.

Techniques inspired from the field of mathematical and structural rigidity theory, and Monte-Carlo based geometric dynamics simulations, have recently gained special attention as they are not affected by time-scale issues and are suitable for high-throughput and big-data style analysis of protein dynamics. These methods can provide very fast predictions of protein flexibility and dynamics and decipher complicated dynamics effects that have been shown to be in agreement with experimental measures of dynamics such as NMR, order parameter measurements, chemical shifts, hydrogen deuterium exchange and others\cite{jacobs2001,wells2005,HenzlerWildman2007,ahmed2011,sljoka2013,zhu2015,kim2017,ye2018,jeliazkov2018,huang2021} and even in structure validation\cite{fowler2020}. Starting with a 3D protein structure, these techniques model protein as a mechanical body-hinge framework\cite{whiteley2005}, whose generic structure is represented as a constrained graph consisting of atoms (nodes) and various connecting edges (covalent bonds, hydrogen bonds, electrostatic interactions, and hydrophobic contacts). Combinatorial graph theoretical algorithms and biophysics techniques are then applied to decompose the protein graph into rigid and flexible regions. Such rigid cluster decomposition of a protein structure can serve as a natural coarse graining step, where hundreds of degrees of freedom are removed from the overall system, and when combined with Monte-Carlo techniques and geometric simulations it provides a low computational complexity alternative to MD simulations for sampling wide regions of high dimensional conformational space. Other extensions of this approach can also be used to generate stereochemically correct transition trajectories between initial structure and end structure. This type of protein simulations is based on simplified physics accounting for only the strongest and the local scale interactions: covalent bond geometry, steric exclusion of atomic spheres, hydrophobic tethers, and local polar interactions including hydrogen bonds together with salt bridges. This set-up allows geometric simulations to be computationally cheap while being sufficiently detailed and informative at individual atomistic level.

\begin{figure}
\centering
\subfigure[]{\label{fig:b}\includegraphics[width=70mm]{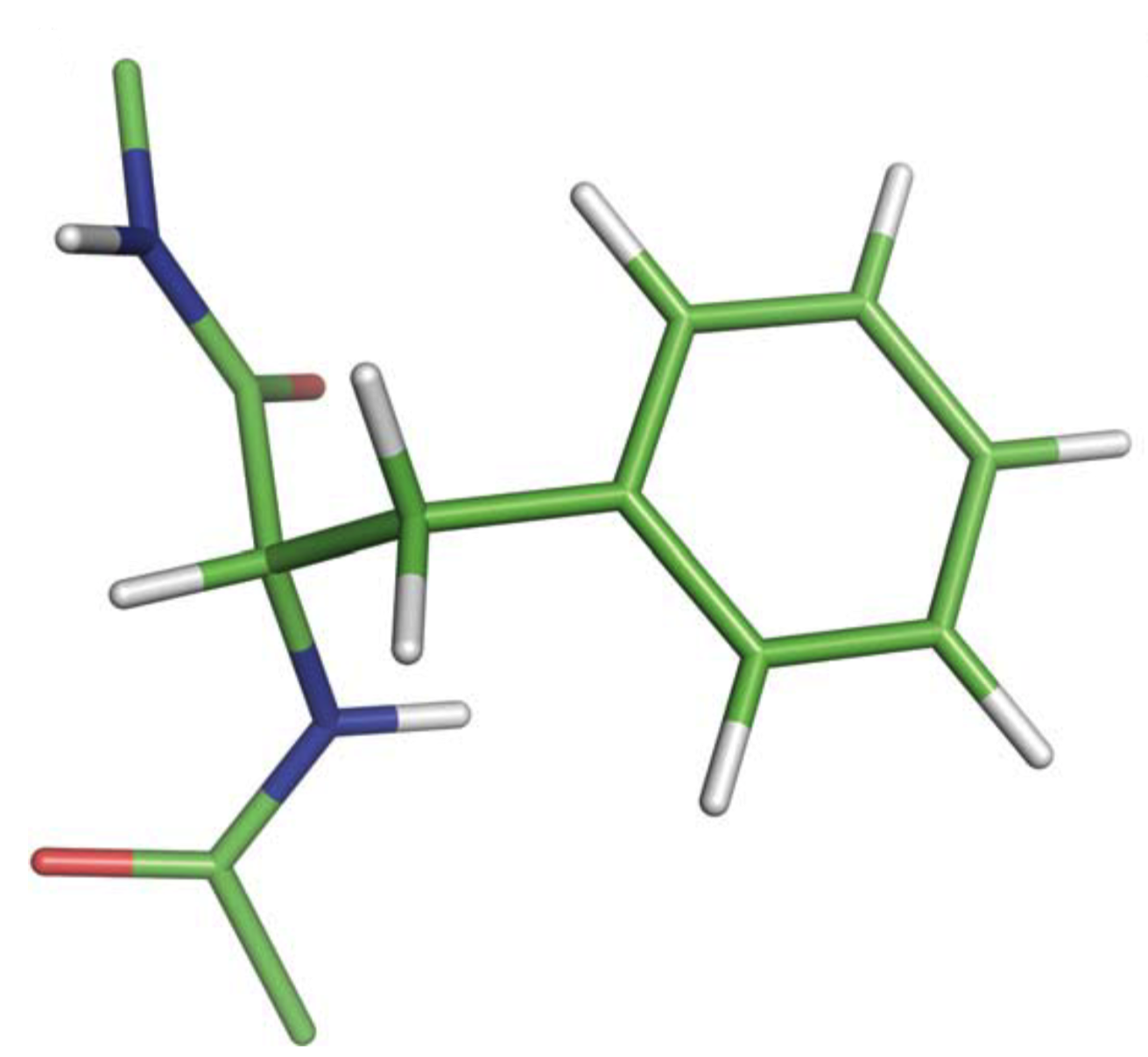}}
\subfigure[]{\label{fig:b}\includegraphics[width=70mm]{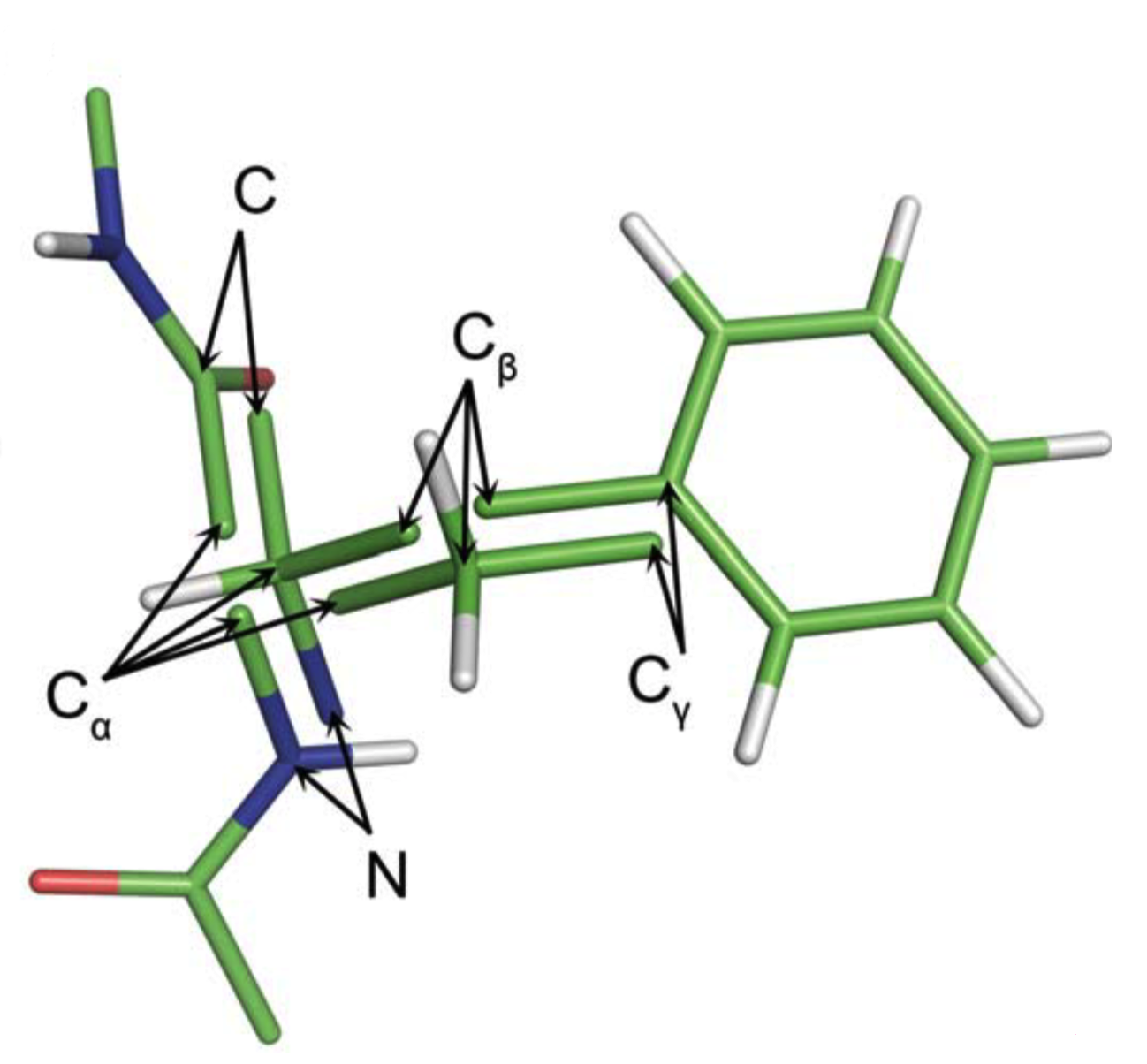}}
\caption{Example of phenylalanine decomposition into rigid units: (a) Stick model with main-chain atoms of neighbouring residues; (b) Atom sets decomposed into rigid units with covalent bond distances and angles locked. In this representation, a single atom may have multiple copies, each belonging to a different rigid unit (pointed out with arrows). Figure is adapted from \cite{farrell2010}}.
\label{fig:0}
\end{figure}

Rigidity theory-based geometric molecular simulation methods FRODAN \cite{farrell2010}, FRODA \cite{wells2005}, including normal mode based geometric simulation approaches NMSim, RCNMA  \cite{ahmed2011} and their extensions, typically utilize Monte Carlo sampling for efficient conformational ensemble generation. The important feature of these methods is that they are an all atom-models, but run approximately $10^4$ times faster than conventional MD simulations \cite{jacobs2012}. The gain in speed comes from underlying rigidity theory concepts that they are employing. Rigidity theory is used to decompose protein structure into rigid subunits that are assumed to be undeformable while relative motion between units is permitted. The rigid cluster decomposition of the protein structure leads to a system with a reduced overall number of degrees of freedom. This leads to significant simplification of potentials that are used for simulations and results in substantial speed up in performance. Despite these complexity reductions, these methods provide results that correlate well with experimental measures of dynamics and classical MD simulations \cite{wells2005,seyler2015,jolley2008}.


In comparison to other geometric simulation methods, FRODAN is optimized to produce greater amplitudes of motion and can handle extremely high flexibility as found in intrinsically disordered proteins \cite{zhu2015}. This is because FRODAN emphasis is put on small templates rather than large scale rigid units as it is used in FRODA. FRODAN explores motions in directions compatible with the constraint network, rather than normal mode exploration as carried out in NMsim. FRODAN is highly suitable to probe dynamics of systems where substantial unfolding and refolding of the protein is present, or where significant conformational changes occur, making it a good approach for analysis of antibody dynamics. 

FRODAN can be run in two modes: targeted and non-targeted. The targeted mode determines the transition pathways between two distinct conformations. It employs biasing force that enables more efficient transitioning between different states. The non-targeted mode explores unbiased random fluctuations starting from a single conformer, which is specially useful for highly flexible proteins as wide conformational space can be sampled \cite{zhu2015}.

In this chapter we focus on application of FRODAN to study various aspects of IgG2a monoclonal antibody dynamics. For this task, FRODAN is run using the non-targeted mode. We utilized the non-targeted mode, rather than the targeted, due to the lack of available clearly segregated states for these proteins. We also briefly consider extensions by combining FRODAN simulations with search algorithms from reinforcement learning to enhance conformational sampling.

\section{Materials and Methods}

Here we give a short introduction in how FRODAN performs dynamics and conformational sampling of protein structures and then offer an extension that uses Monte Carlo Tree Search to guide the conformational search. We have carried out analysis on Protein Data Bank structures 1IGT; 4G6K; 5W6C.
\subsection{FRODAN - Geometric dynamics simulations}



As an initialization step, we first construct a constrained graph representation of the 3D protein structure. In this graph model, vertices represent a set of all protein atoms including hydrogens, while edges represent covalent and non-covalent interactions. To capture  specifics of covalent bond geometry, each amino acid of the protein is divided into rigid subcomponents based on rigid bond distances and rigid 3-body angles. Dihedral angles for single covalent bonds are not constrained, whereas dihedral angles for higher bond orders are treated as locked. The grouping of atoms into rigid clusters is done using rigidity theory method FIRST \cite{jacobs2001}. Figure \ref{fig:0} illustrates an example of how the amino acid phenylalanine is decomposed into rigid units. Within 20 standard amino acids, the largest rigid unit is the planar indole group in tryptophan. The smallest rigid units consist of three atoms, such as the C-OH in the side chain of tyrosine or serine. Each rigid unit of the system has six conformational degrees of freedom (i.e. three translational and three rotational). The rigid units contain "embedded atoms", whose positions depend entirely on the degrees of freedom of their corresponding rigid unit. Several "embedded atoms" may correspond to the same physical atom, for example $C_\alpha$ and $C_\beta$ atoms in Figure \ref{fig:0} (b). The physical position of these atoms is set to be at the mean of their "embedded atoms".

In order to define the allowed and disallowed conformations between rigid units, various geometric constraints between units are imposed. Non-bonded pairs of atoms must not violate a specified distance constraint value. Depending on the atom types involved, the value may differ. These cut-off distances have been calibrated using MD. To maintain backbone dihedral angles within the allowed regions in the Ramachandran plot, additional minimum distance constraint between certain pairs of main-chain atoms are imposed. To keep side-chain torsion angles in correct configurations there are constraints between 1-4 bonded atom pairs when atoms 2 and 3 are single-bonded and each is tetrahedrally coordinated. Hydrogen bonds and hydrophobic contacts are preserved by maximum-distance constraints between pairs of interacting atoms. Hydrogen bond is kept if it passes an energy score defined by the cutoff value using Mayo potential \cite{jacobs2001}.

FRODAN conformational sampling in non-targeted mode begins with the atoms in their initial positions from the input structure. Atoms are constrained according to the geometry of initial structure. Random perturbations at each step of the sampling are applied on the system according to the following procedure: 1) Randomly perturb identified rigid units of the system; 2) Enforce the set of geometric constraints; 3) All residues with a severe constraint violations are returned back to their original positions; 4) Re-enforce the set of constraints for the updated system; 5) Fit the current state to the initial structure used at the beginning of the simulation; 6) Calculate net change relatively to the initial structure used at the beginning of the current step, if it passes the treshold save it and use it as initial for the following step.


\subsection{Geometric simulations and search algorithms}

In the non-targeted mode, FRODAN has the potential to sample wide regions of conformational space. However, a common objective is the exploration of conformational states that have some desired characteristics. For instance, transition pathways that traverse the space and reach conformations of interest from a starting state are often important to understand. Running FRODAN in a non-targeted mode alone may be too slow for these applications. If desired start and end conformers have experimentally solved structures, a reasonable choice is to use FRODAN in a targeted mode instead. Targeted simulations can quickly and accurately find transition pathways between two conformations \cite{farrell2010}. Since the conformations of interest often lack either starting or final crystal structures, FRODAN's targeting in this case is less applicable. To overcome this difficulty, FRODAN non-targeted mode can be combined with search algorithms, for example Monte Carlo Tree Search (MCTS) - a type of reinforcement learning algorithm that has demonstrated impressive performance at various classes of tasks including protein folding \cite{browne2012,shin2019}. The strength of MCTS is the ability to balance between exploration and exploitation when search is expanded. This feature allows MCTS to quickly find candidates with desired metrics and at the same time explore wide conformational space.  MCTS can be viewed as an expanding search tree where edges are individual FRODAN runs and nodes are the best conformers from prior runs. The selection of the next node for the expansion (used as a starting conformation for the FRODAN run) is based on the so called Upper Confidence Bound (UCB) score defined as:


\begin{equation} \label{eq:1}
UCB_i=X_i+C\sqrt{\frac{2\log N}{n_i}}
\end{equation}

\noindent where $X_i$ is a quality metric of the conformer that needs to be maximized, $C$ is a constant, $N$ is a number of times the parent node has been visited, $n_i$ is a number of times $i$th child has been visited. The first term of the equation facilitates exploration of candidates with high quality metrics, while the second one facilitates exploration of the less visited nodes. Contribution of both terms is regulated by the constant $C$. To additionally facilitate diversity of sampling, the first term of the (\ref{eq:1}) can be multiplied by the factor $\alpha^{N_{sim}}$, where $\alpha$ is a constant and $N_{sim}$ is a number of the similar conformers already found. Each node of the tree has its $UCB_i$ score. Every time the new node is added, tree's $UCB_i$ scores are updated. The node with the highest $UCB_i$ score will be expanded next.

\subsection{FRODAN simulation procedure}
Here we describe the basic procedure for performing geometric dynamics simulations with FRODAN.

FRODAN simulation commences with a PDB file, which usually requires preprocessing: addition of hydrogen atoms, removal of waters and selection of alternate sidechain conformations. In this chapter, we perform simulations on three PDB files with ids: 1IGT; 4G6K; 5W6C. 1IGT corresponds to the crystal structure of the whole IgG2a monoclonal antibody. 4G6K and 5W6C correspond to the isolated Fab domain with H3 lengths 10 and 15, respectively. After removing waters and single-atom groups, we added hydrogens using MolProbity server \cite{molprobitywebsite}.

Next, the list of options defining how the FRODAN non-targeted run will be conducted must be specified. This list must contain hydrogen energy cutoff value, number of steps for the run and threshold for the new conformer output. The hydrogen energy cutoff value (in kcal/mol) parameter specifies hydrogen bonds that will be included in the simulation run. In the example where whole antibody complex is considered, energy cut-off is set 0 kcal/mol. In the single FAB domain structures, selected energy cut-offs are 0 and -2 kcal/mol.

The number of steps for the run defines the total number of steps that will be made during the simulation run.  At each simulation step the whole system is perturbed, this is followed by an enforcement of constraints. If the new generated conformation differs from the previous more than a specified treshold value, it will be outputted and used as the initial conformational in the next step. This will be continued the defined number of times. In the examples in this chapter, the number of outputted conformers per individual run was set 30000.

The new conformer output parameter specifies the output frequency. The system's conformation will be outputted if it has moved away from the previous conformation more than the set threshhold value. For the considered examples in this chapter, the treshold value was set to 1 {\AA}.

Once the options above are specified FRODAN run can be initiated. The run will continue until the termination criterion is reached.

\begin{figure}
\centering
\subfigure[Overall antibody structure.]{\label{fig:b}\includegraphics[width=78mm]{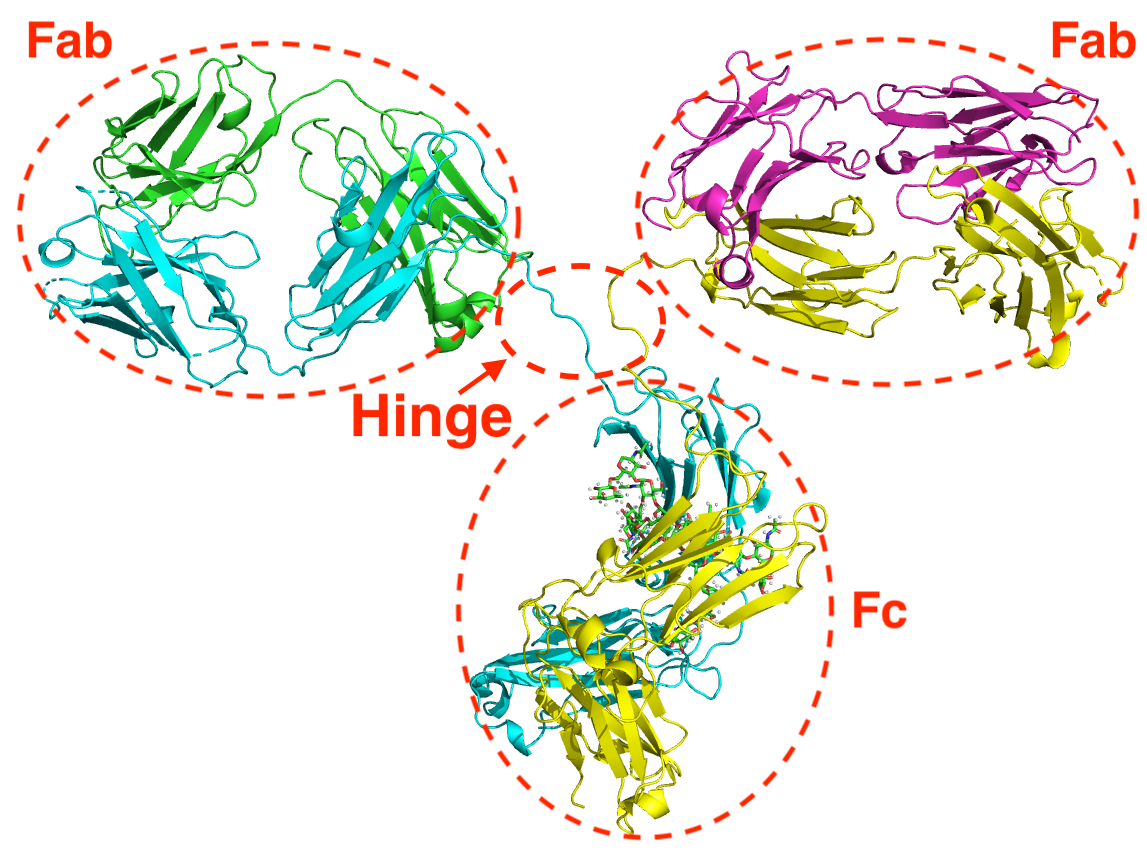}}
\subfigure[Fab arm with highlighted CD3 H3 loop.]{\label{fig:b}\includegraphics[width=50mm]{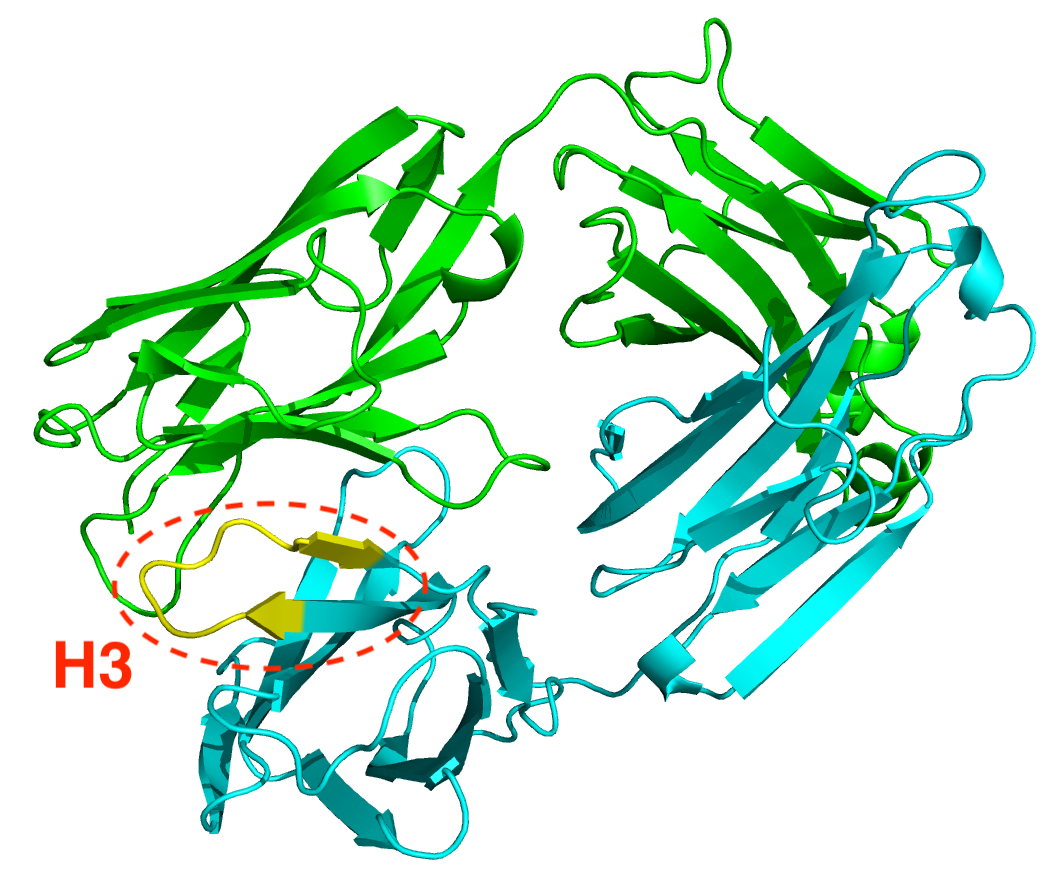}}
\caption{The essential antibody domains and architecture.}
\label{fig:01}
\end{figure}

\begin{figure}
\centering
\subfigure[]{\label{fig:b}\includegraphics[width=38mm]{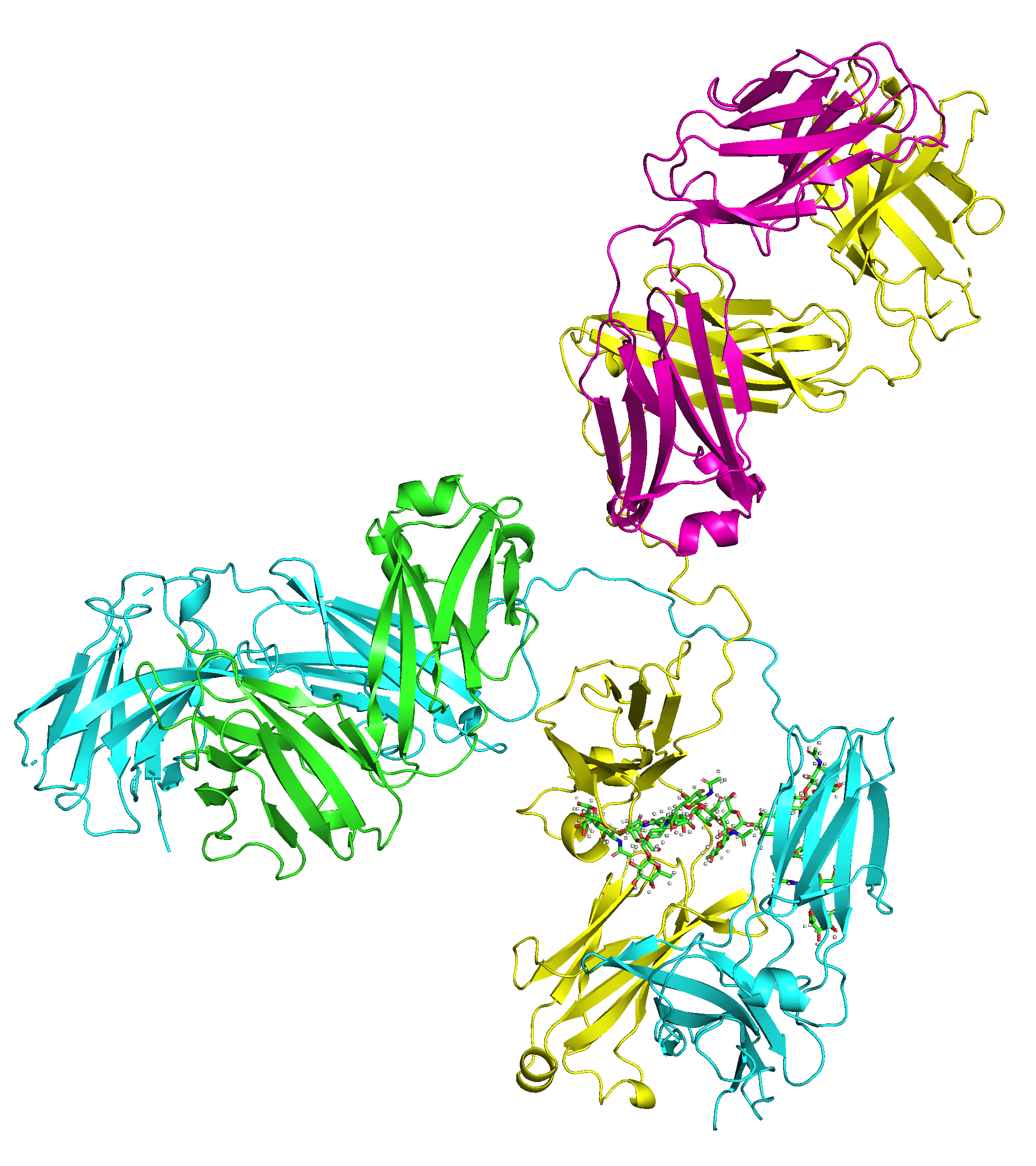}}
\subfigure[]{\label{fig:b}\includegraphics[width=30mm]{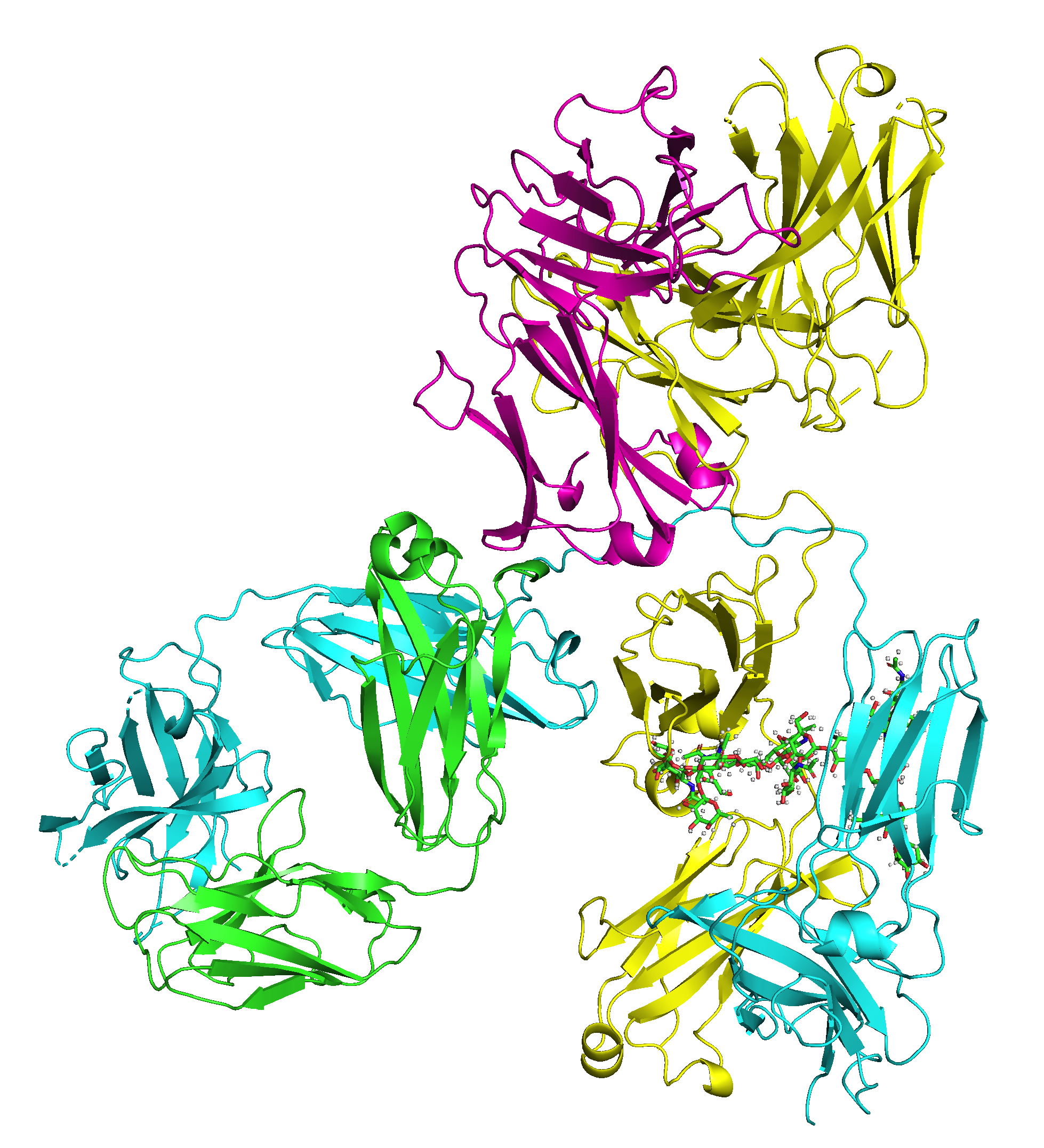}}
\subfigure[]{\label{fig:b}\includegraphics[width=34mm]{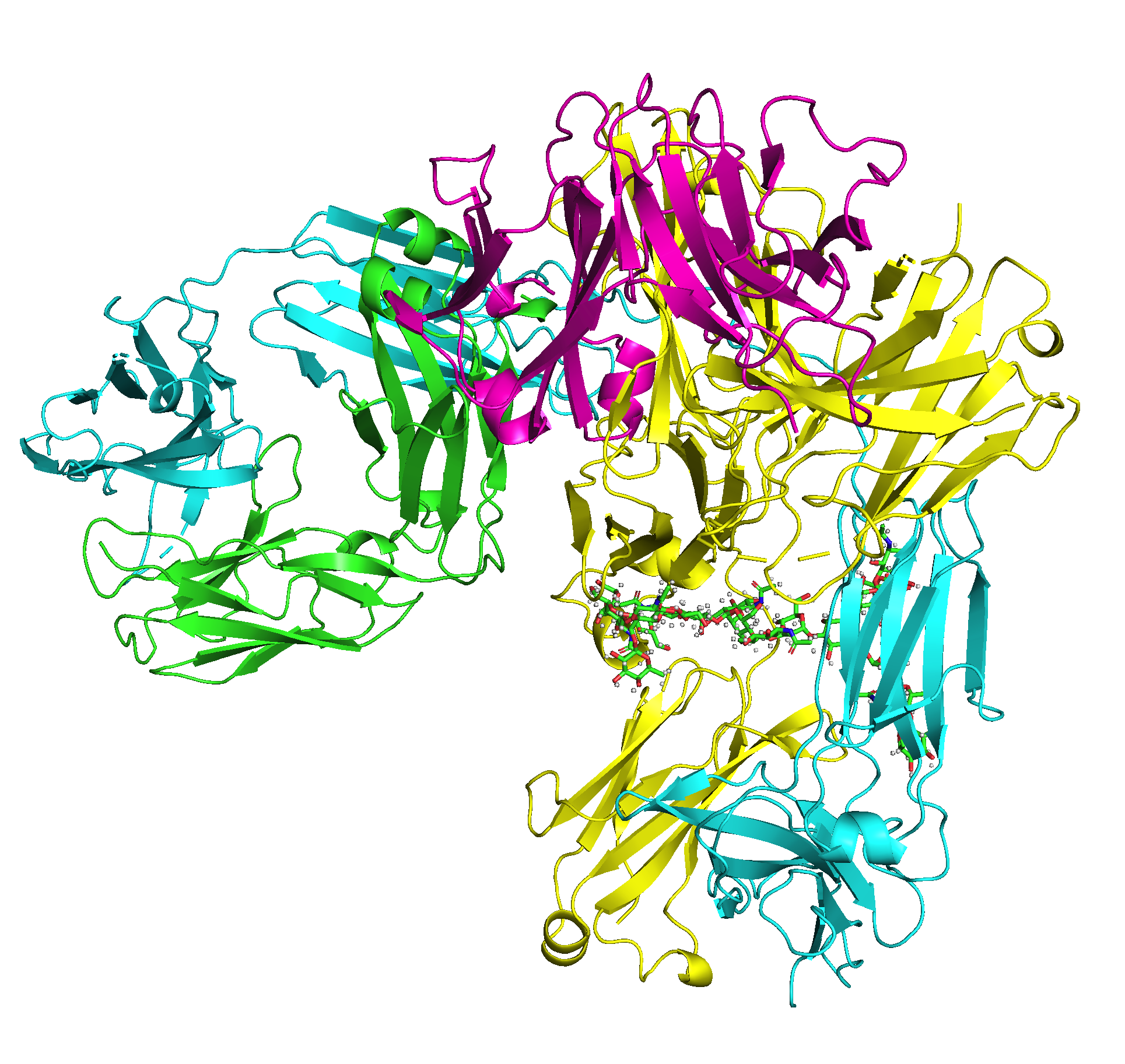}}\\
\subfigure[]{\label{fig:b}\includegraphics[width=34mm]{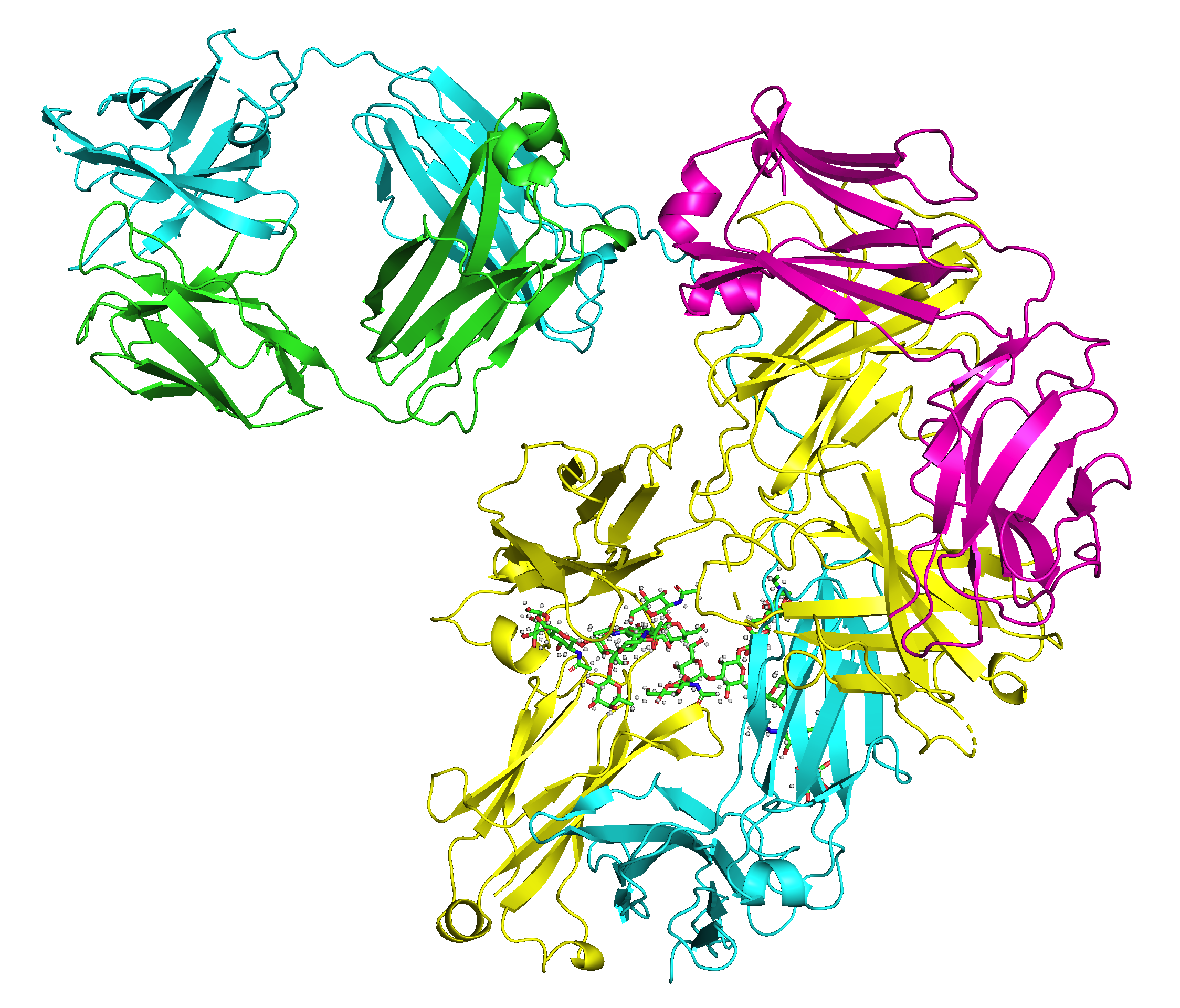}}
\subfigure[]{\label{fig:b}\includegraphics[width=34mm]{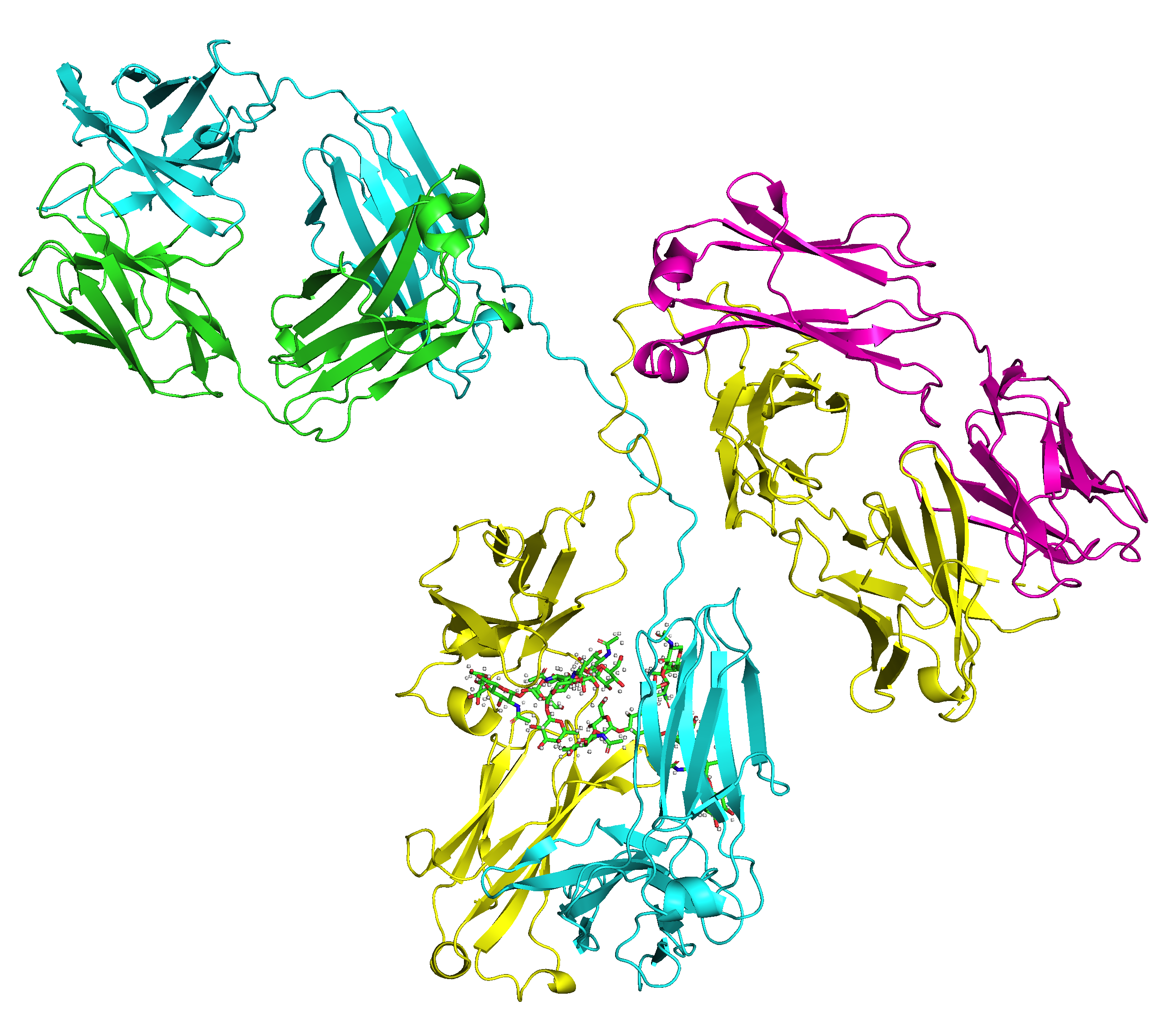}}
\subfigure[]{\label{fig:b}\includegraphics[width=28mm]{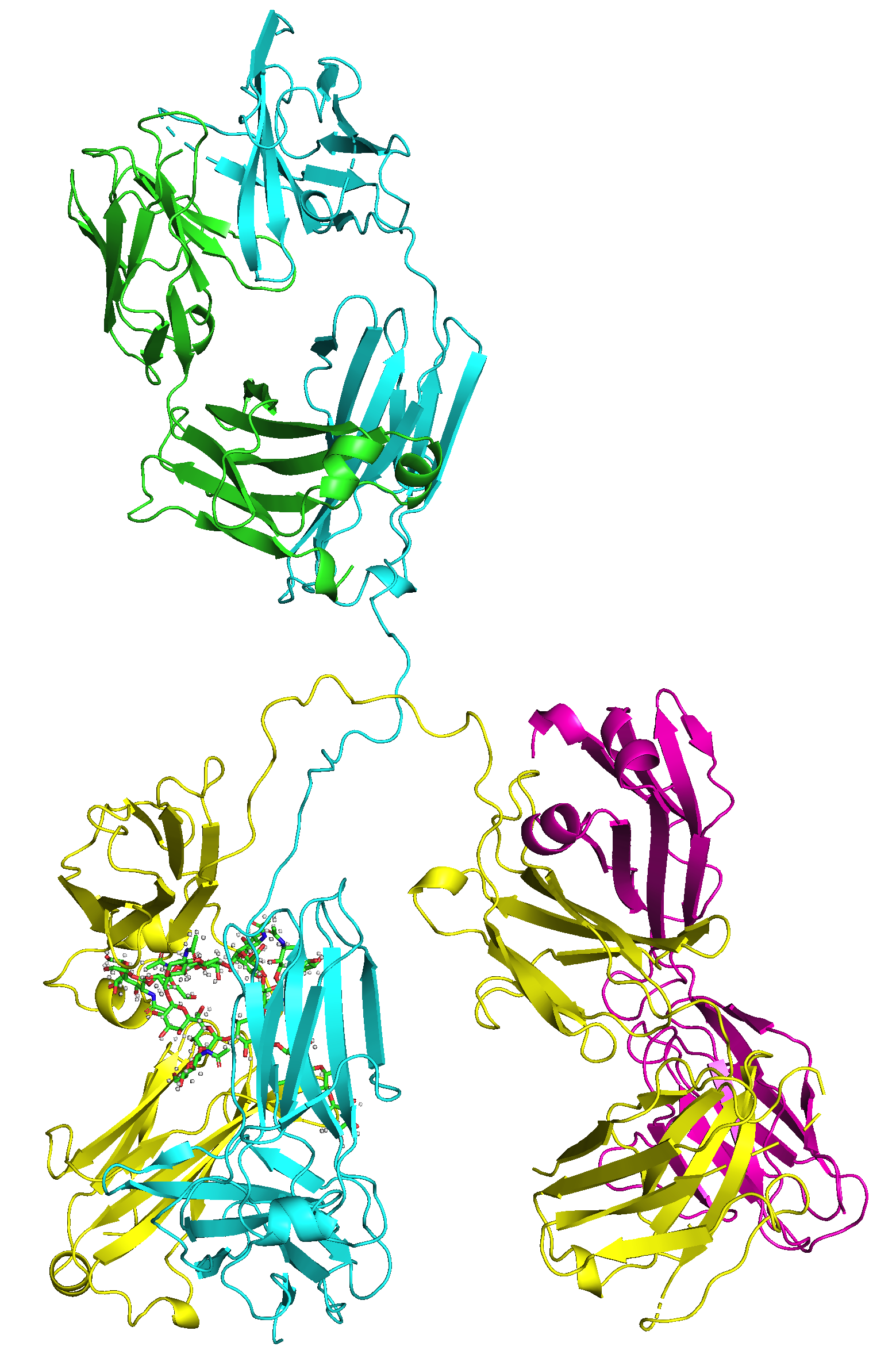}}
\caption{The transition between representative conformational states of Fab arms obtained using FRODAN non-targeted run. The transition is viewed from (a) to (f).}
\label{fig:1}
\end{figure}

In this chapter we also use FRODAN in combination with MCTS in order to perform more efficient transition of IgG2a monoclonal antibody between two distinct Fab arm conformations, see Figure \ref{fig:01}. The initial state is shown in Figure \ref{fig:1} (a) whereas the target state is shown Figure \ref{fig:1} (f). For FRODAN guidance, MCTS employs backbone RMSD between generated conformer and the target. Our objective is to find conformer(s) with the lowest RMSD value. We use the same MCTS implementation as Shin et al. \cite{shin2019} with MD simulation replaced by FRODAN. FRODAN is run in non-targeted mode. At every expansion step 1000 conformers are generated. At the end of expansion step, conformer with the lowest RMSD relative to the target structure is selected and if it is lower than conformer's RMSD used at the beginning of the current run the candidate node is added to the search tree. Afterwards $UCB_i$ of each node in the updated tree are recalculated according to the equation (\ref{eq:1}). The RMSD score (the $X_i$ in the equation (\ref{eq:1})) is multiplied by -1 in order to give expansion preference to nodes with lower RMSD values. The search tree is expanded predefined number of iterations. For the example in this chapter, 400 iterations is set. Other MCTS parameter values used here are the default ones from Shin et al. \cite{shin2019}.

\section{Applications - Antibody Dynamics}

\begin{figure}
\centering
\subfigure[]{\label{fig:b}\includegraphics[width=40mm]{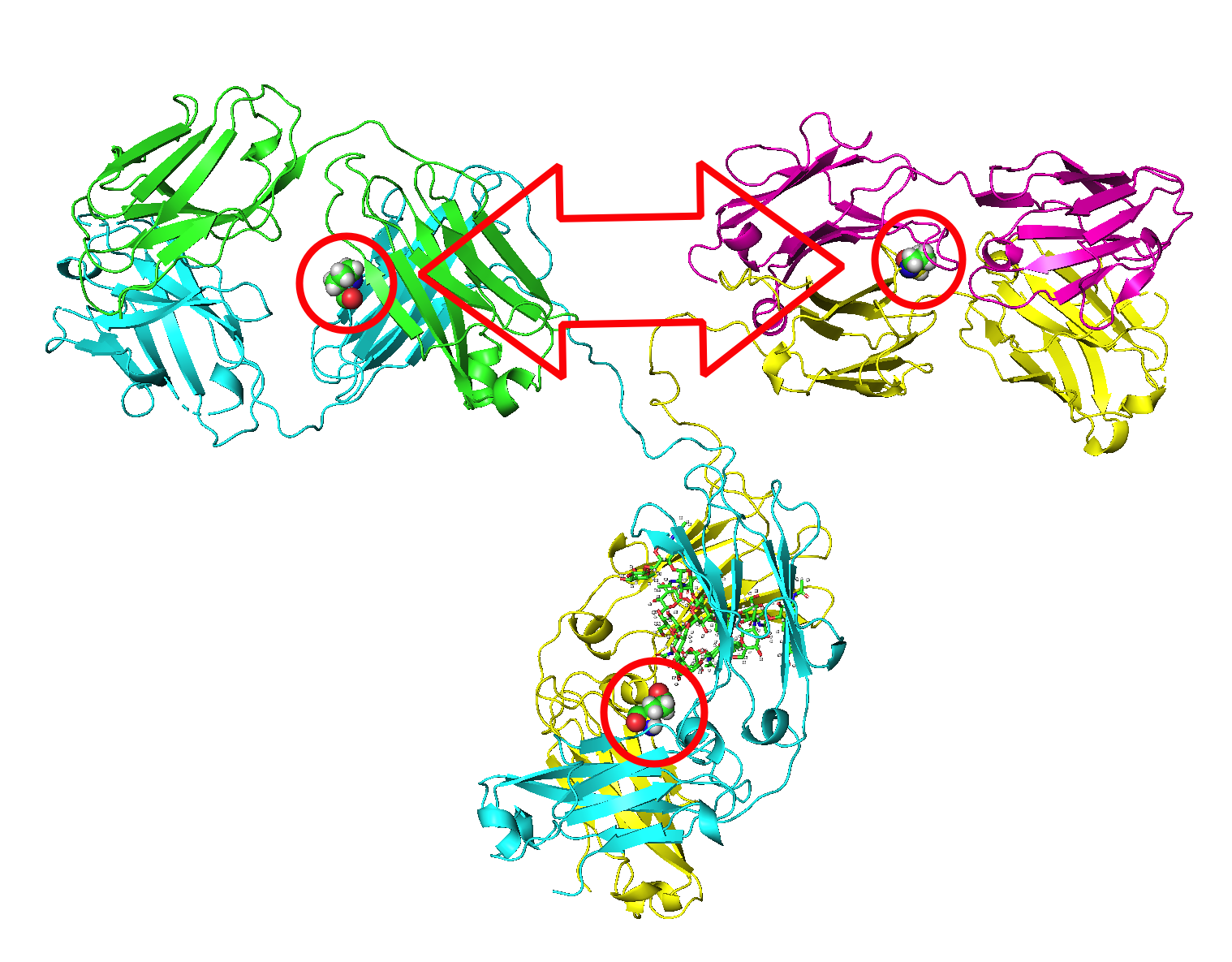}}
\subfigure{\label{fig:b}\includegraphics[width=75mm]{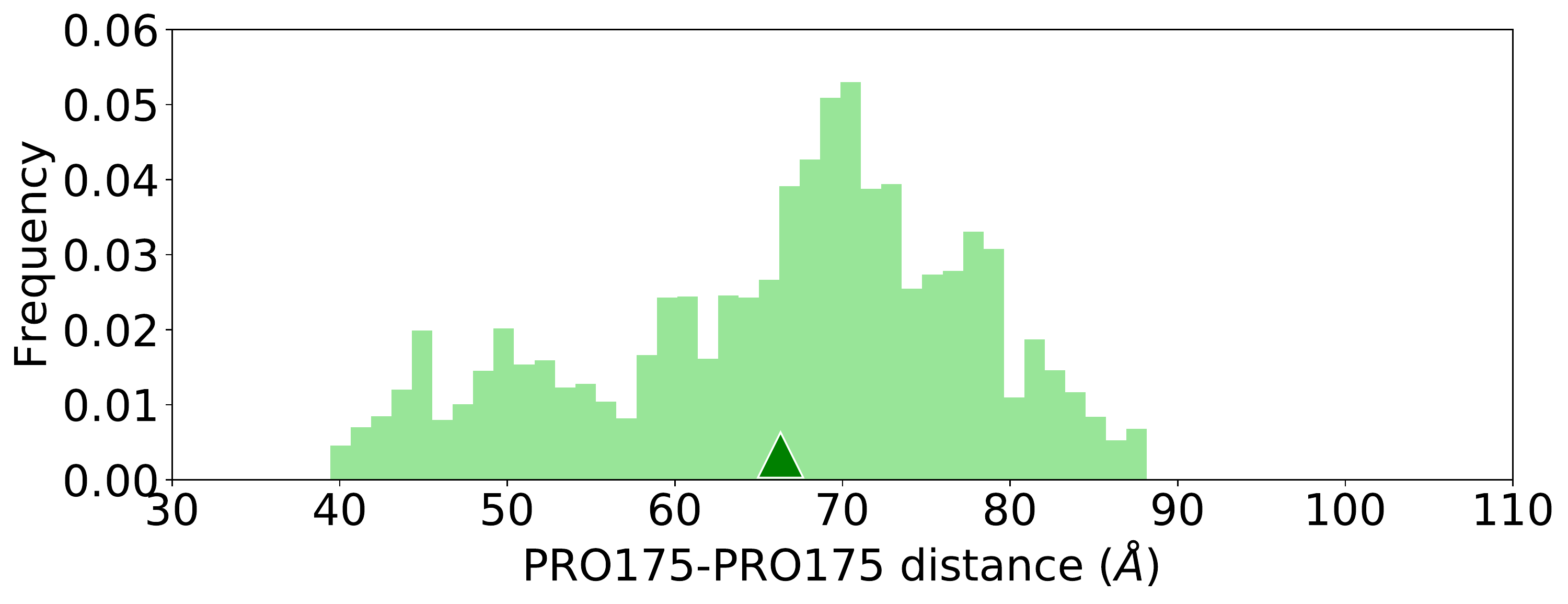}}\\
\subfigure[]{\label{fig:b}\includegraphics[width=40mm]{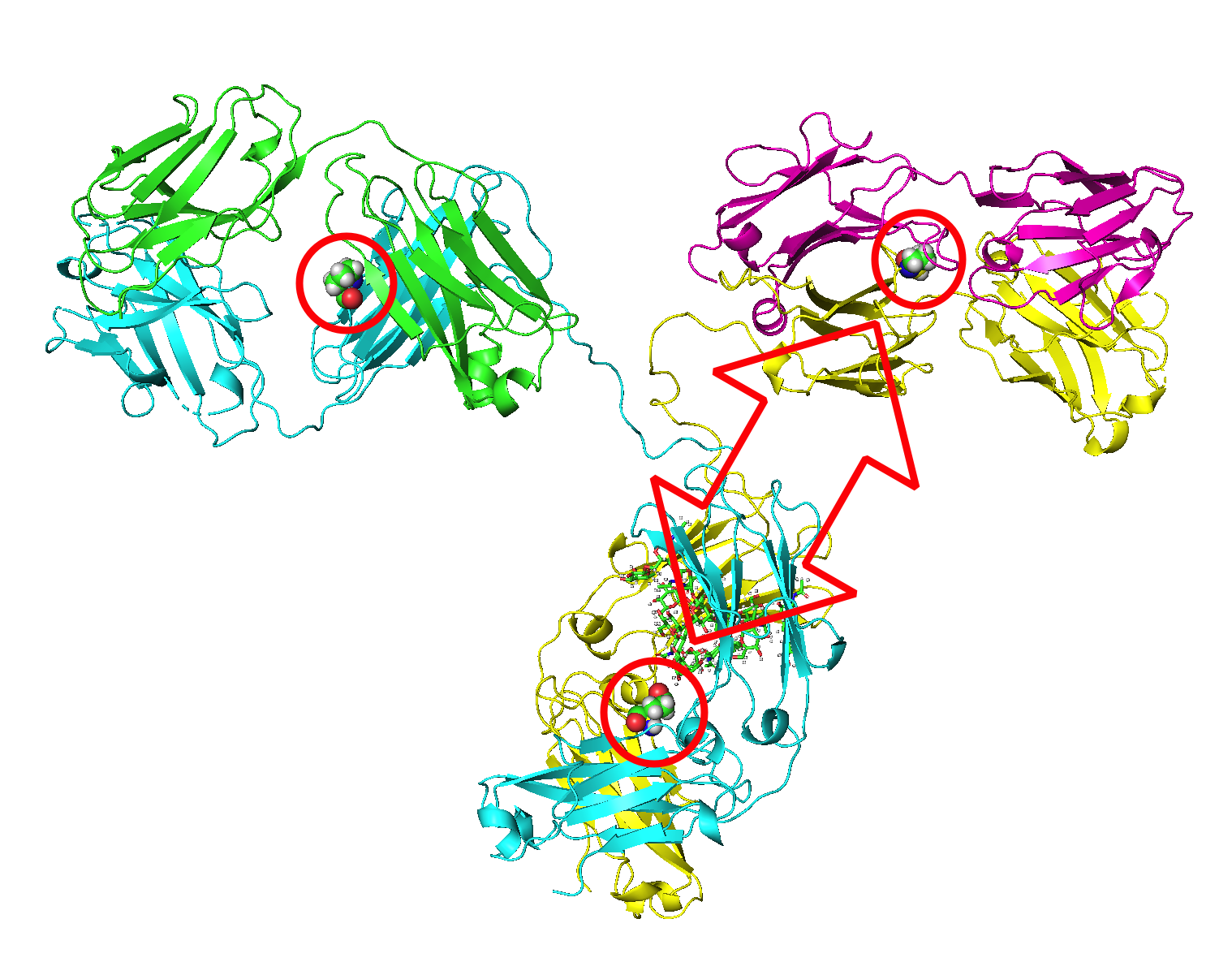}}
\subfigure{\label{fig:b}\includegraphics[width=75mm]{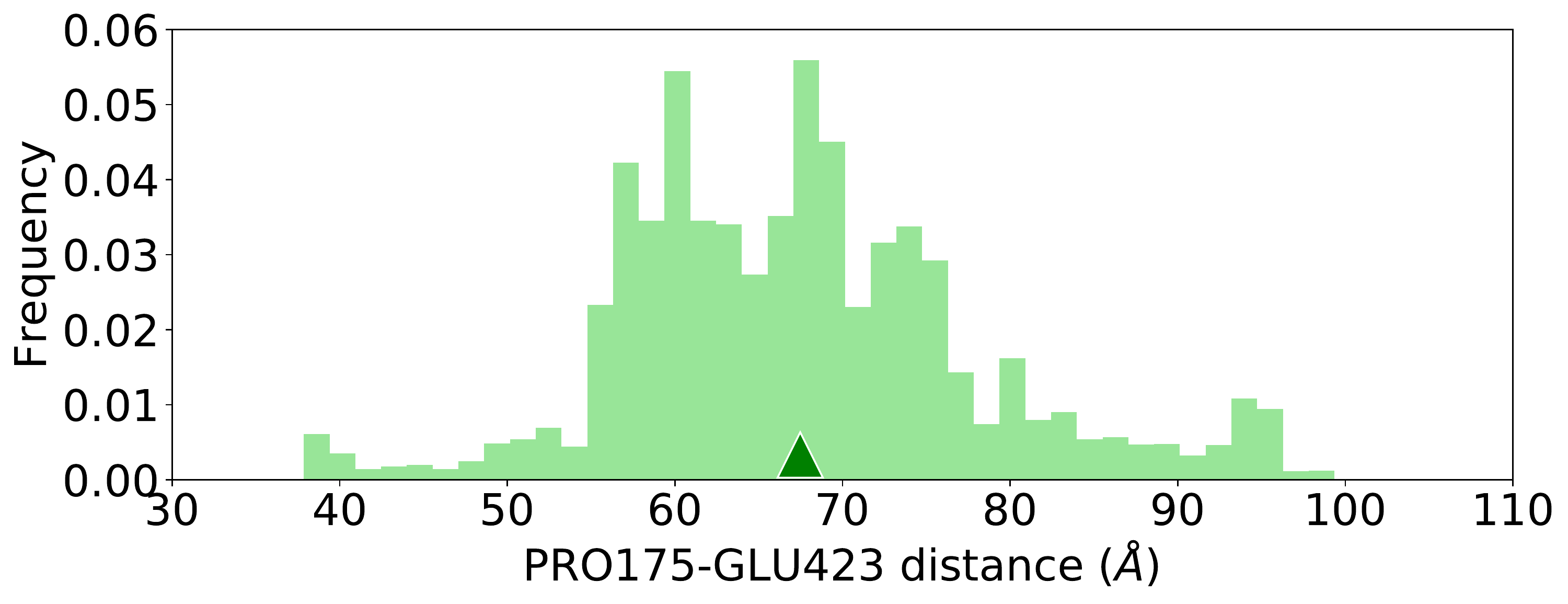}}\\
\subfigure[]{\label{fig:b}\includegraphics[width=40mm]{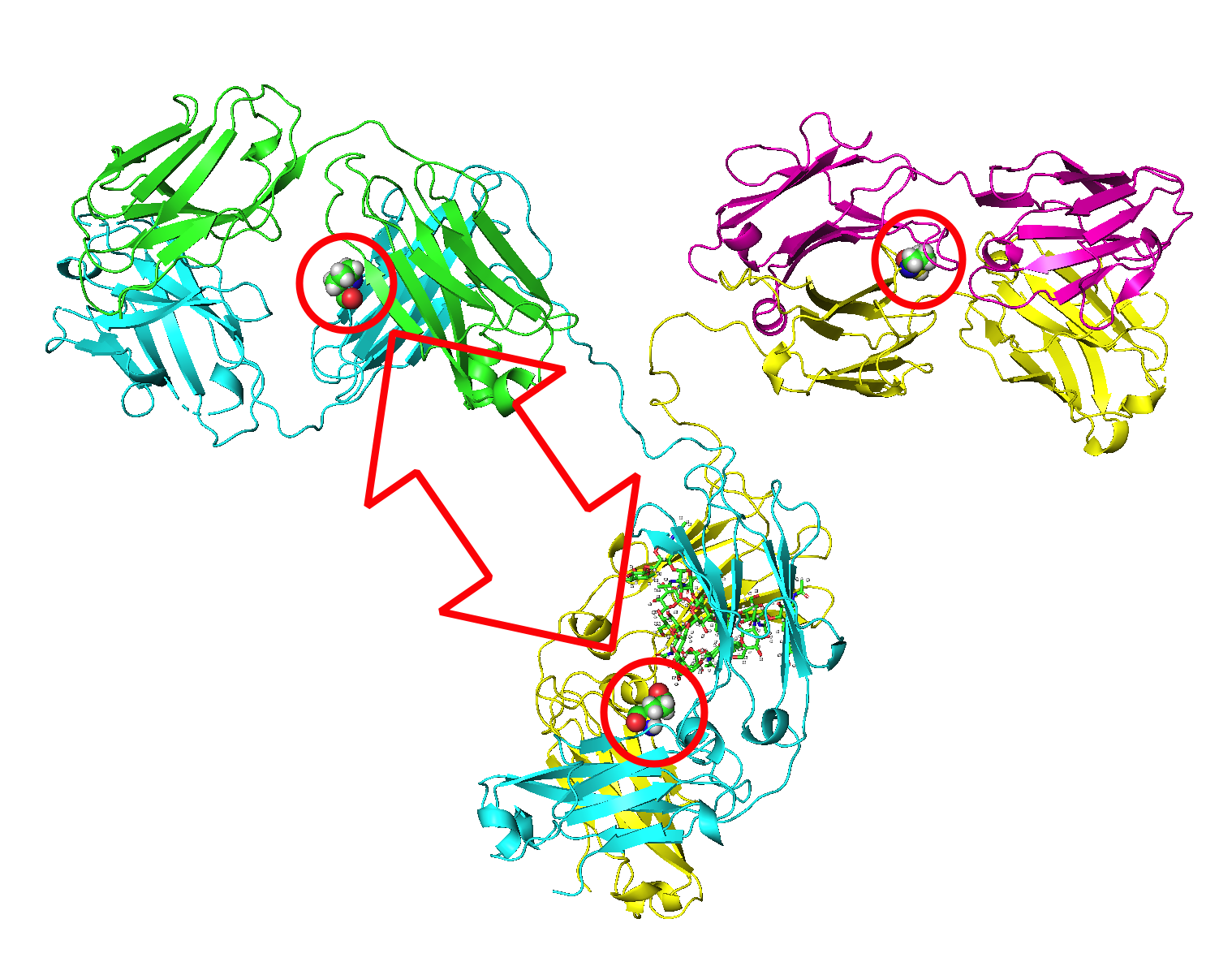}}
\subfigure{\label{fig:b}\includegraphics[width=75mm]{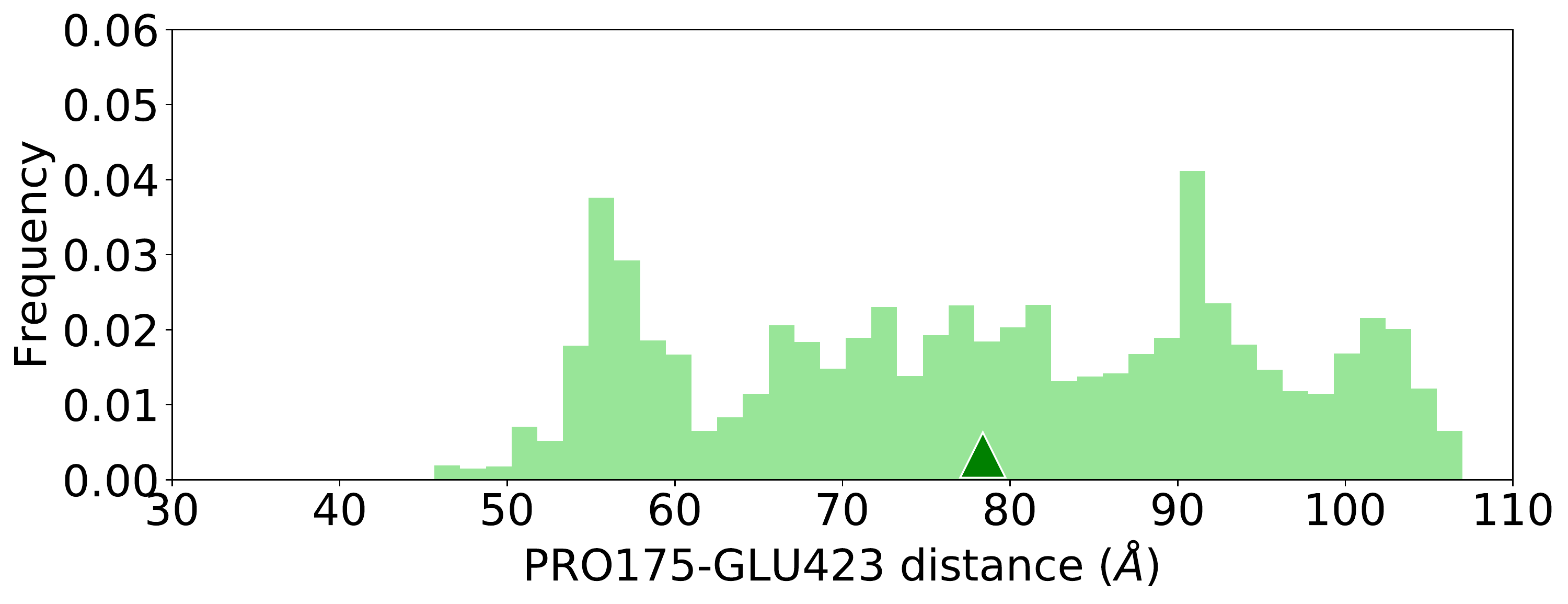}}
\caption{Distributions of domain distances: (a) Fab-Fab; (b) Fab-Fc; (c) Fab-Fc.}
\label{fig:2}
\end{figure}

Selected representative transition snapshots between two distinct Fab arm conformations generated by FRODAN during the non-targeted run for the whole IgG2a monoclonal antibody using pdb:1IGT as a starting structure are shown in Figure \ref{fig:1}. All three domains of antibody: two Fabs and Fc, exhibit relative degree of domain-domain flexibility and dynamics. Fab arms exhibit large amplitude motions relative to each other and to the Fc domain, consisting of rotational, translational and screw motions of rigid domain-domain motions \cite{gerstein1994,flores2006}. Additionally, there exist small and medium scale flexible motions, not identified here. These observations are in agreement with previous experimental observations \cite{bongini2004,zhang2015}. These experimental studies captured similar conformations to the ones generated by FRODAN. In order to quantitatively validate sampled dynamics, distance variations between representative domains were calculated, see Figure \ref{fig:2}. Sampled distributions for Fab-Fab, and both Fab-Fc pairs overlap reasonably well with distributions reported in \cite{bongini2004,zhang2015}. These results indicate the geometric simulations methodology capacity to explore relevant conformational states of the antibody together with transitions between these states. Similar information can be extracted using sophisticated experimental techniques like cryo-electron tomography \cite{bongini2004} or by fitting selected snapshot to the experimentally derived ones from the ensemble of antibodies \cite{zhang2015}. In the example provided here, essential motions together with conformational ensemble can be extracted from a single FRODAN run.

\begin{figure}
\centering
\subfigure[]{\label{fig:b}\includegraphics[width=60mm]{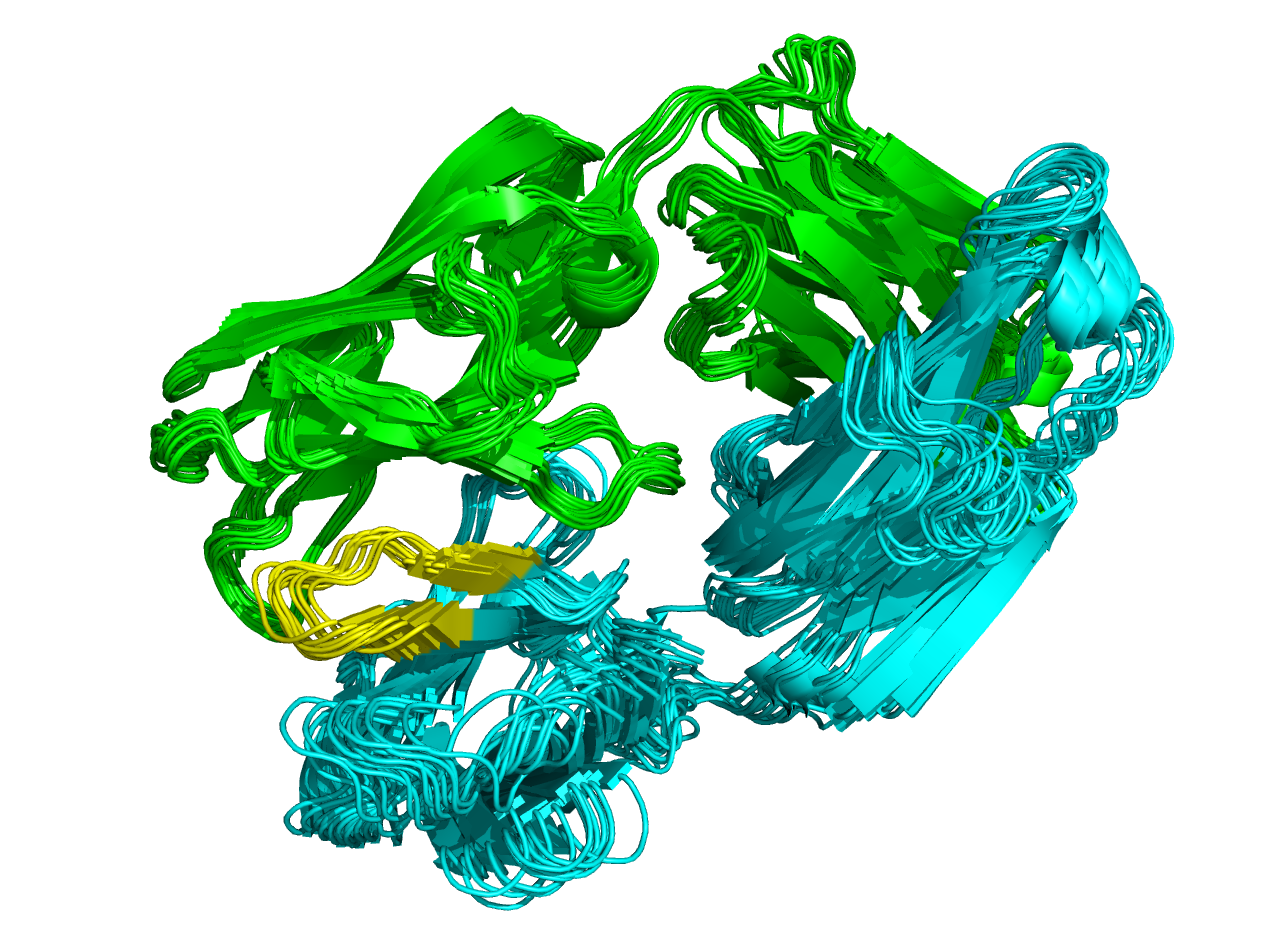}}
\subfigure[]{\label{fig:b}\includegraphics[width=56mm]{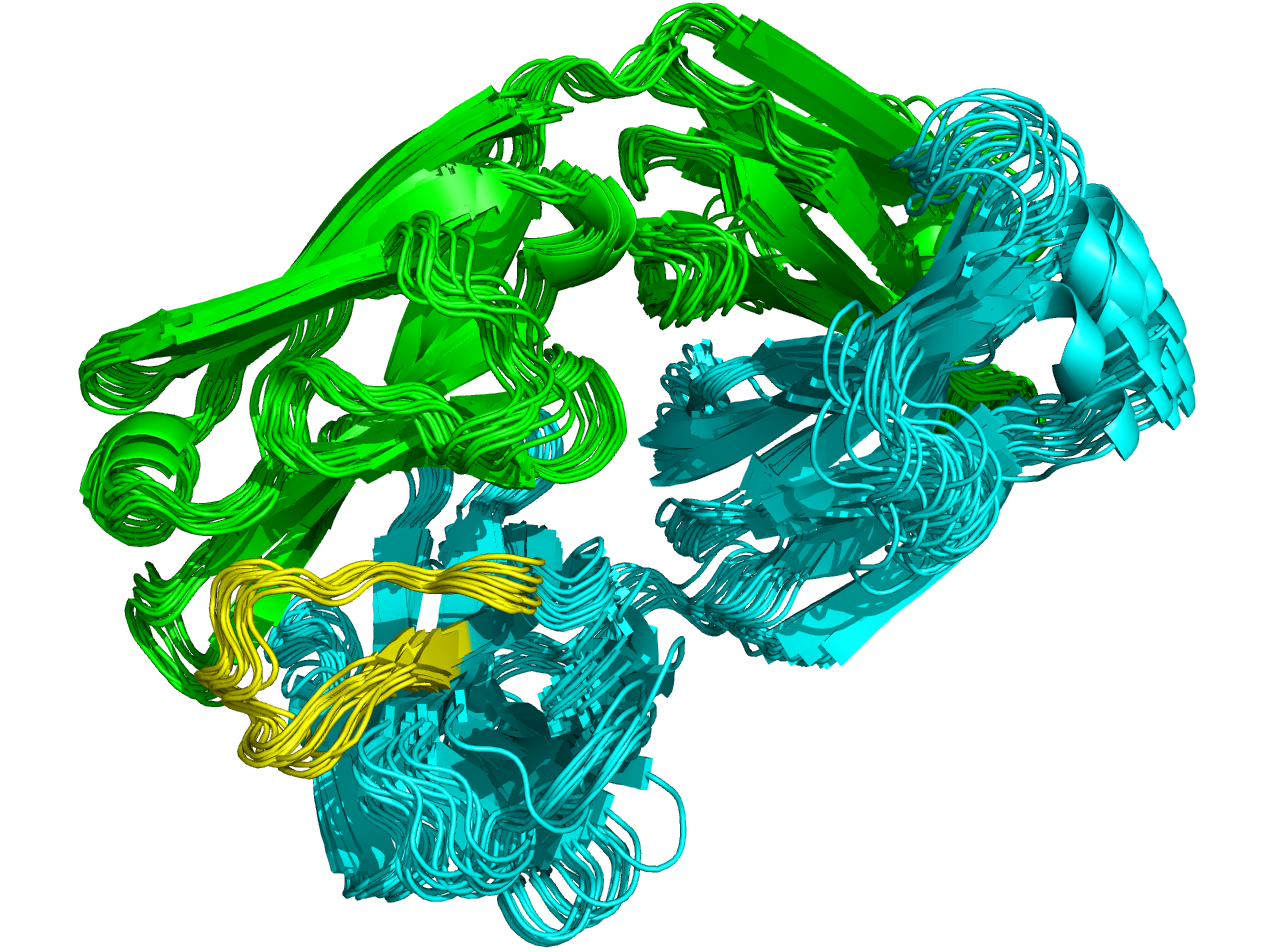}}\\
\subfigure[]{\label{fig:b}\includegraphics[width=54mm]{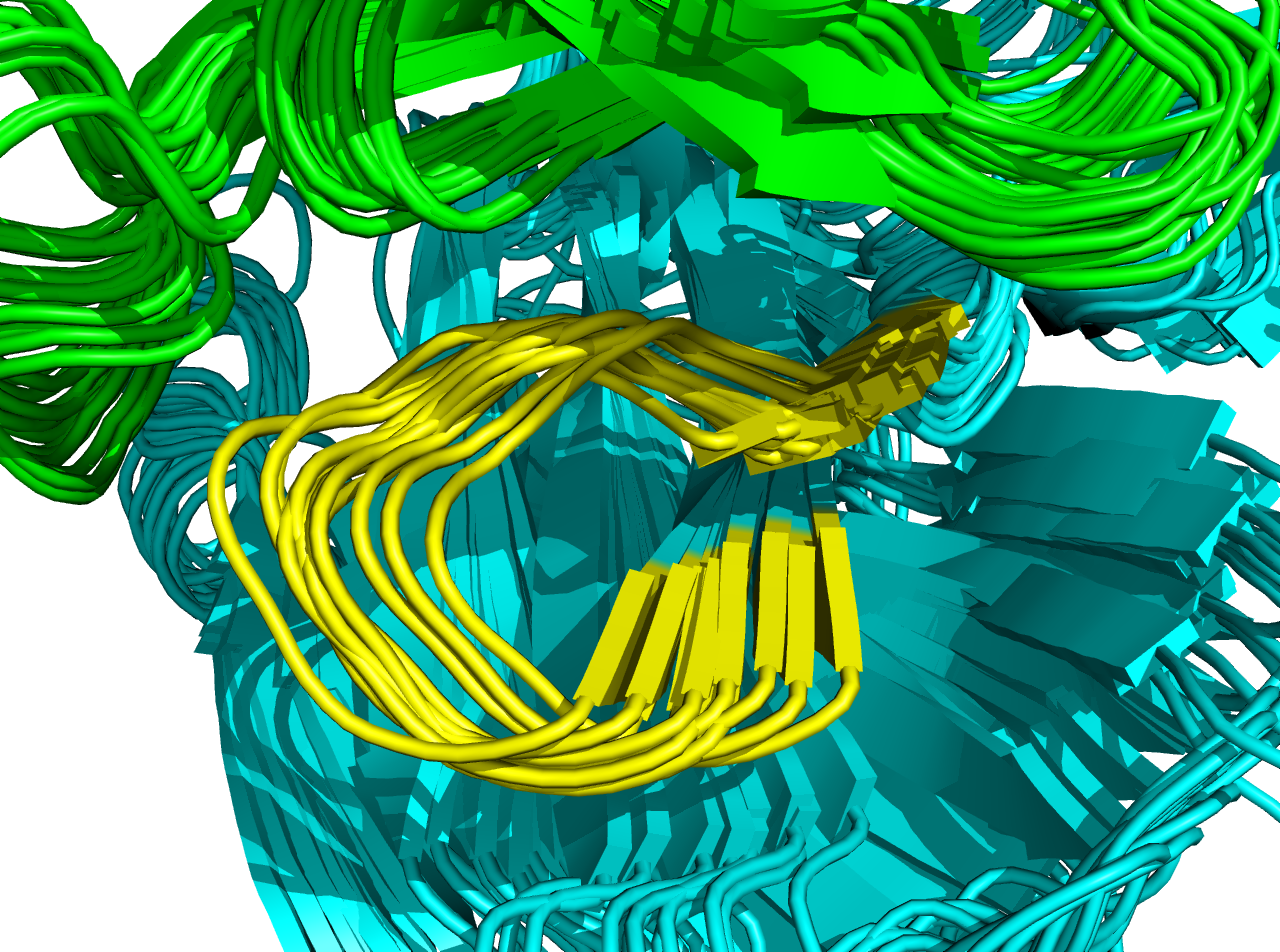}}
\subfigure[]{\label{fig:b}\includegraphics[width=54mm]{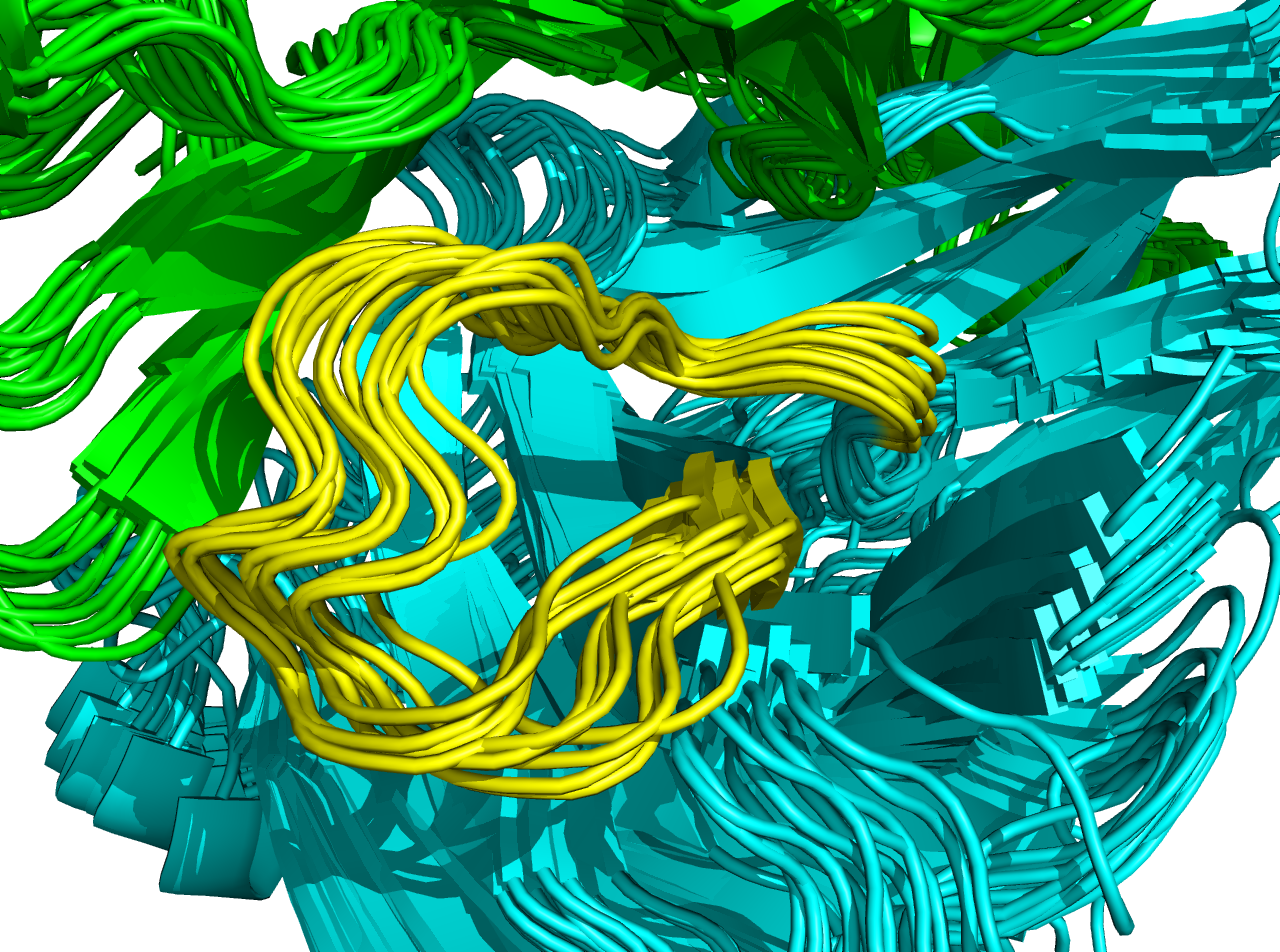}}
\caption{H3 loop (in yellow) dynamics at 0kcal/mol: (a),(c) 4G6K; (b),(d) 5W6C.}
\label{fig:3}
\end{figure}

\begin{figure}
\centering
\subfigure[]{\label{fig:b}\includegraphics[width=60mm]{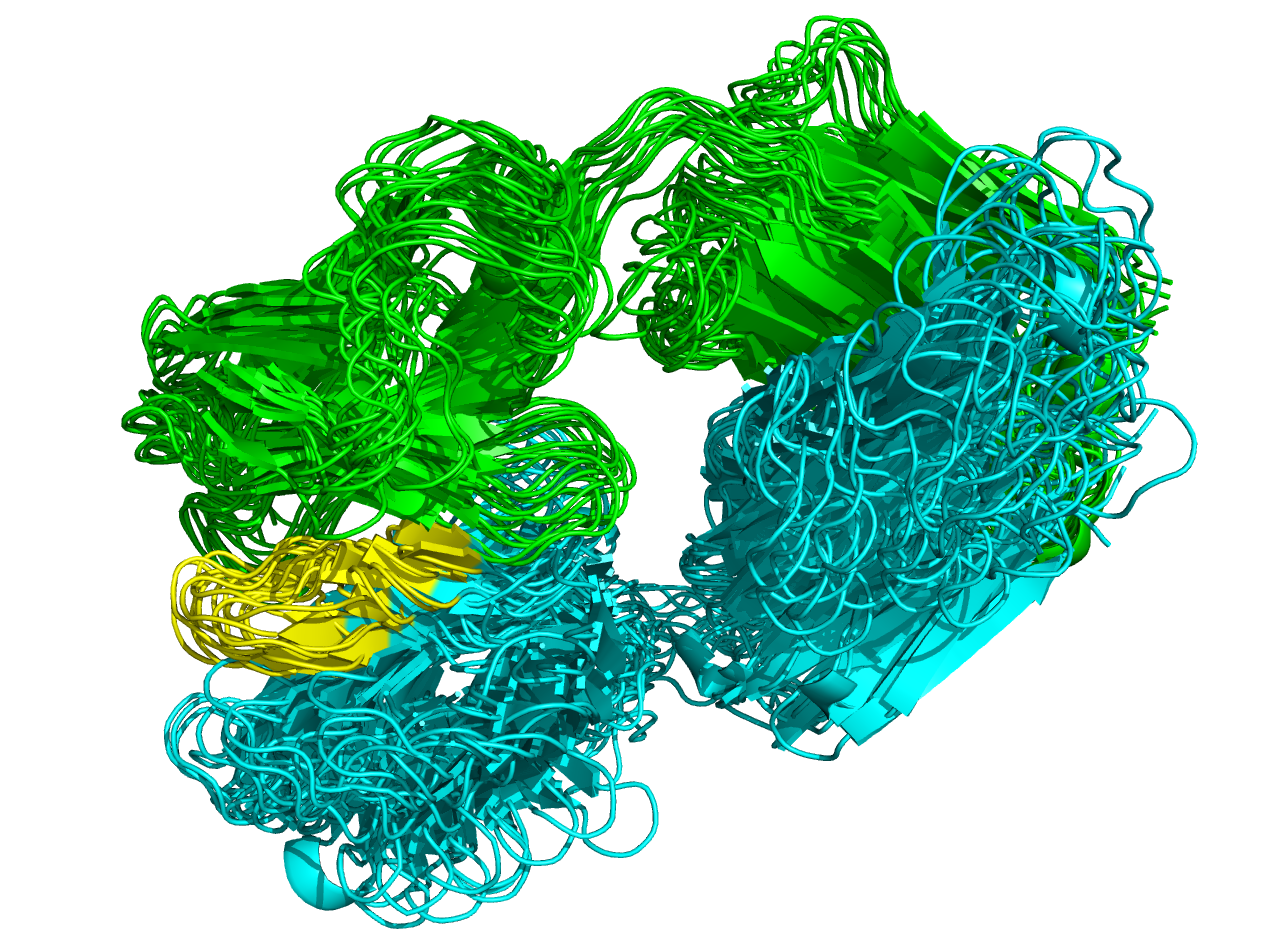}}
\subfigure[]{\label{fig:b}\includegraphics[width=56mm]{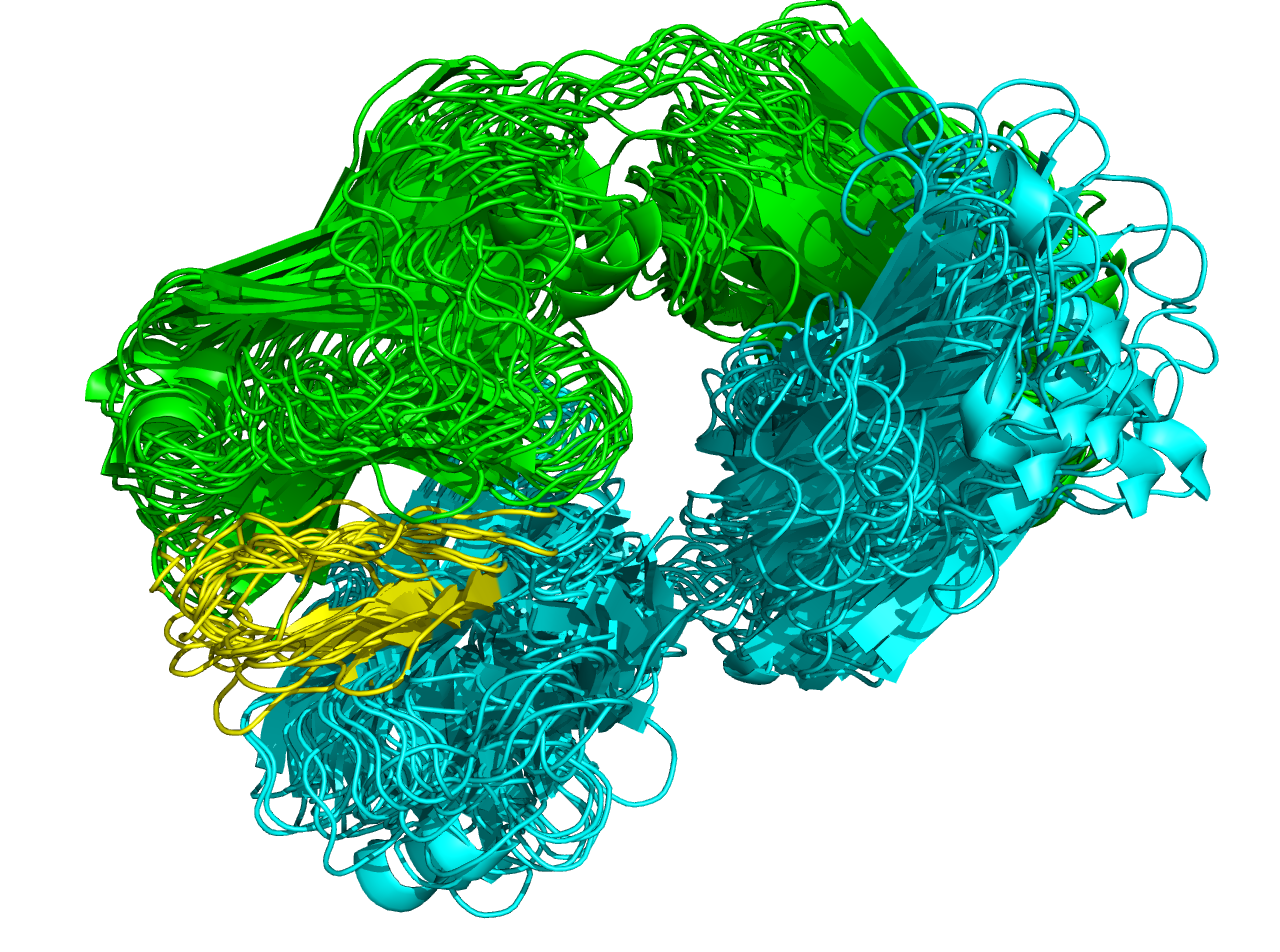}}\\
\subfigure[]{\label{fig:b}\includegraphics[width=54mm]{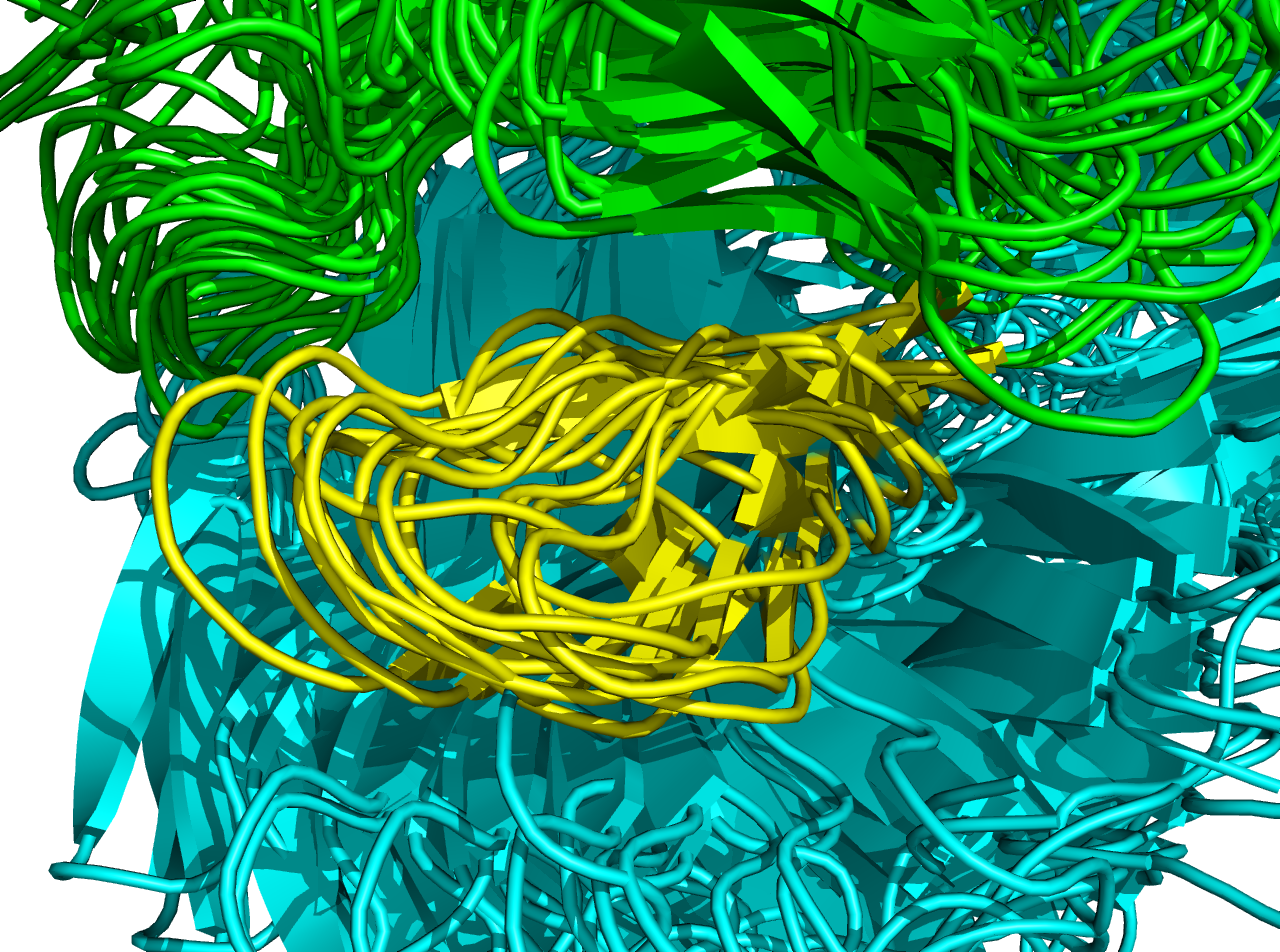}}
\subfigure[]{\label{fig:b}\includegraphics[width=54mm]{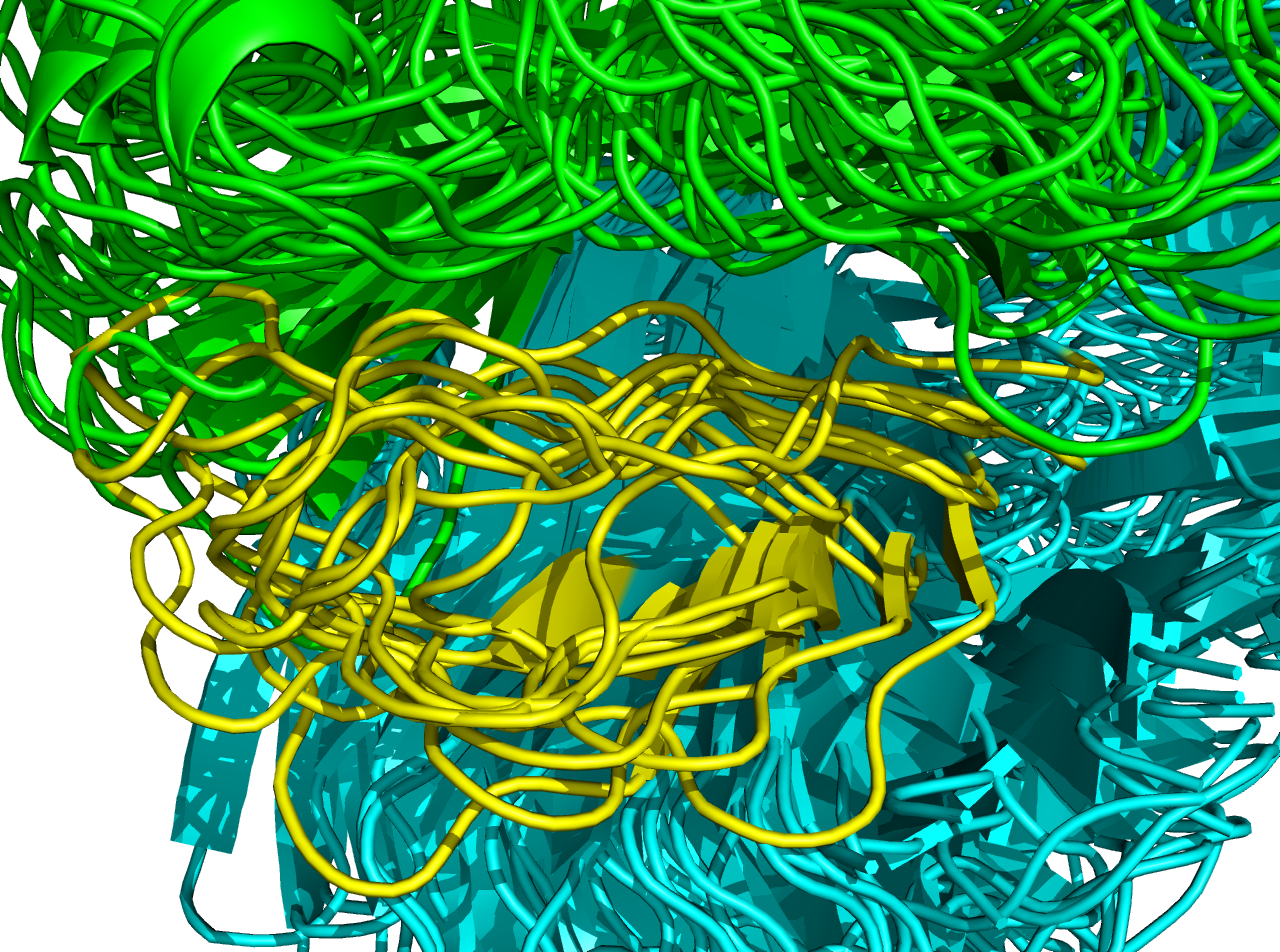}}
\caption{H3 loop (in yellow) dynamics at -2kcal/mol: (a),(c) 4G6K; (b),(d) 5W6C.}
\label{fig:4}
\end{figure}

The representative snapshots generated by FRODAN for isolated Fab domain (4G6K and 5W6C) are superimposed in Figure \ref{fig:3} and \ref{fig:4}. At 0 kcal/mol, the motion of H3 loop in both structures appear to be mainly rigid body (translations, rotations) motions (i.e. lacking internal degrees of freedom). When energy cut-off is reduced to -2 kcal/mol, the motion of the whole Fab domain intensify;  H3 loop starts to show more pronounced internal flexible dynamics (i.e. backbone dihedral angle variations). Particularly, in the case of H3 of length 15, the loop exhibits larger amplitude motions than H3 of length 10. At 0 kcal/mol the difference between the two is mostly indistinguishable.

\begin{figure}
\centering
\subfigure[]{\label{fig:b}\includegraphics[width=54mm]{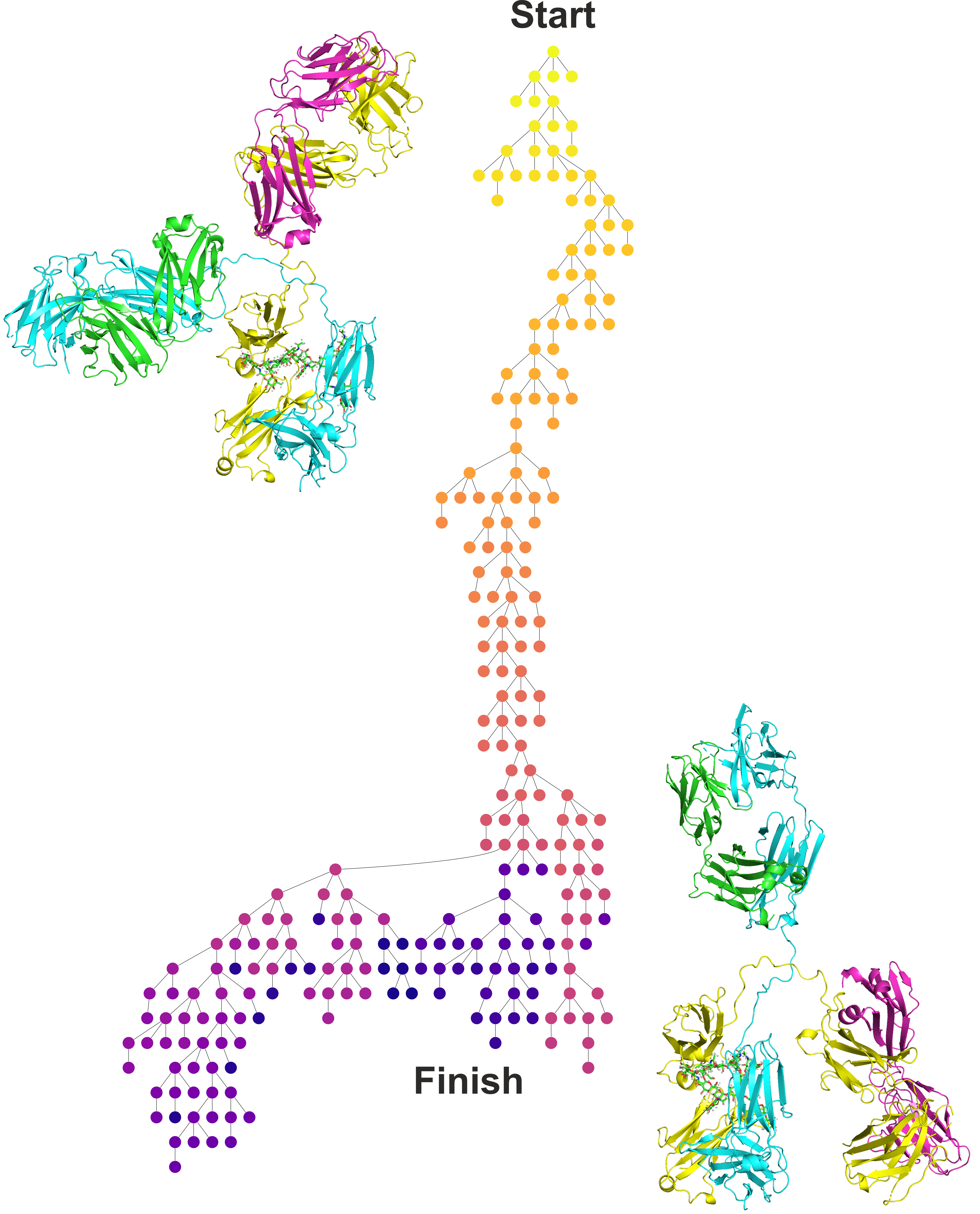}}
\subfigure[]{\label{fig:b}\includegraphics[width=94mm]{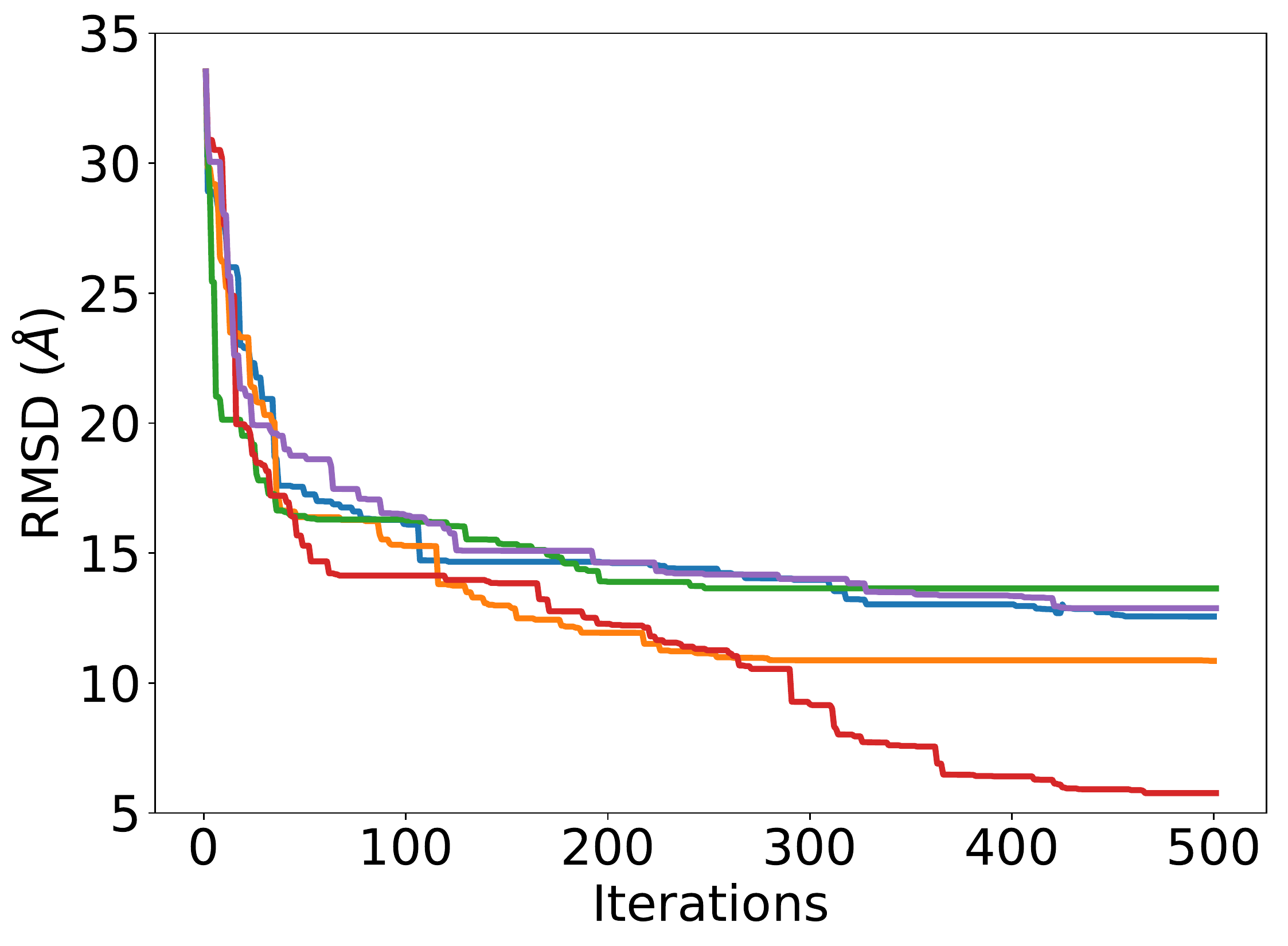}}\\
\subfigure[]{\label{fig:b}\includegraphics[width=54mm]{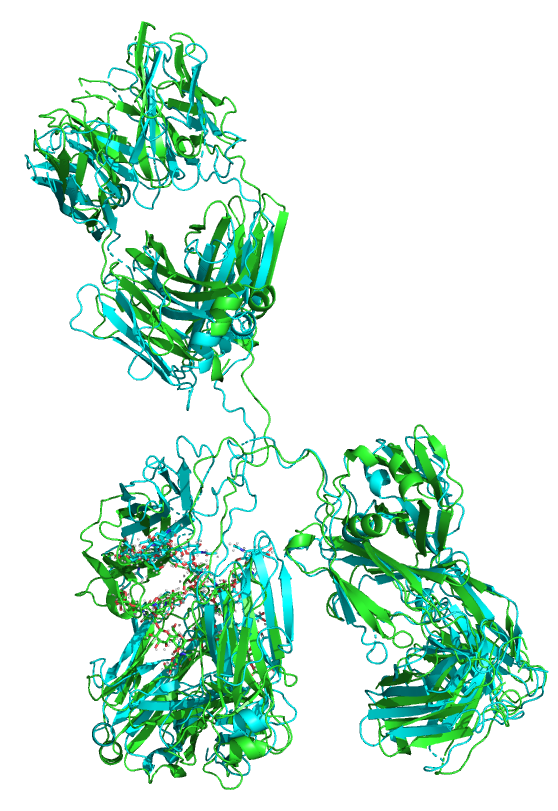}}
\caption{FRODAN combined with MCTS: (a) The search tree example when transitioning from one distinct Fab arm conformation to another; (b) Evolution of RMSD from target during five independent runs. The lowest RMSD found corresponds to 5.8{\AA} red line); (c) The best conformer found (cyan) superimposed on the target (green).}
\label{fig:5}
\end{figure}

\begin{figure}
\centering
\hspace*{-1.0cm}
\subfigure[]{\label{fig:b}\includegraphics[width=48mm]{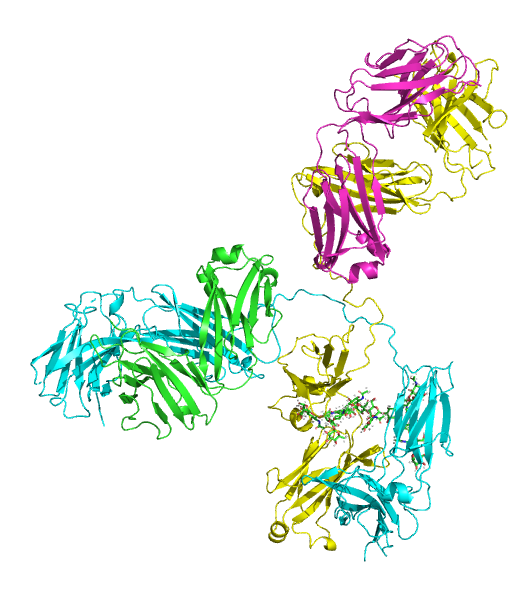}}
\subfigure[]{\label{fig:b}\includegraphics[width=52mm]{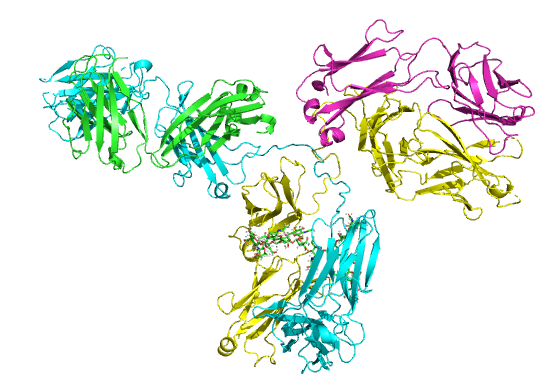}}
\subfigure[]{\label{fig:b}\includegraphics[width=52mm]{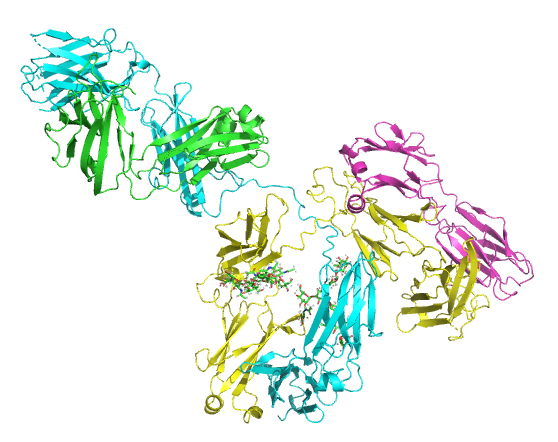}}\\
\hspace*{-1.0cm}
\subfigure[]{\label{fig:b}\includegraphics[width=46mm]{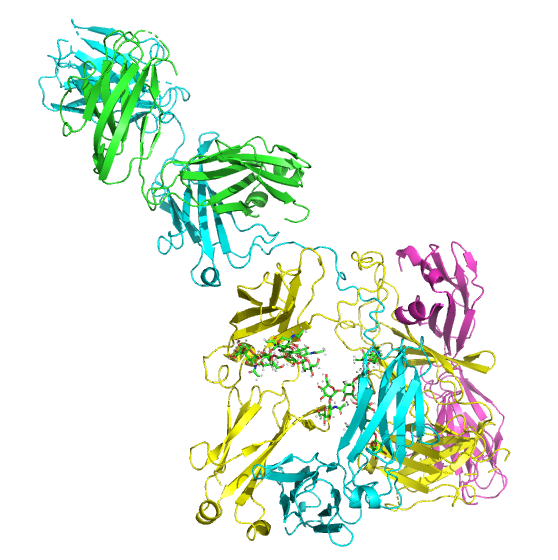}}
\subfigure[]{\label{fig:b}\includegraphics[width=42mm]{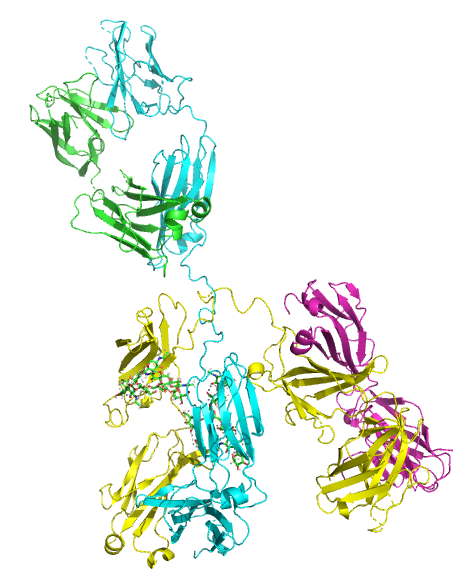}}
\caption{The transition between representative conformational states of Fab arms obtained using FRODAN non-targeted mode combined with MCTS. The transition is viewed from (a) to (e).}
\label{fig:6}
\end{figure}

Results obtained using a combination of FRODAN and MCTS are summarized in Figure \ref{fig:5}. A resulting search tree is shown in Figure \ref{fig:5} (a). The search propagation direction is indicated by color gradient. Nodes that are added at the initial stage of the search are colored in light colors, whereas nodes that are added at the final stages are in darker colors. Figure \ref{fig:5} (b) shows the search progress in terms of the RMSD relative to the target conformation. Five independent MCTS runs are indicated by different colors in the Figure. The most successful run is in red starting from RMSD=33.5{\AA} and reaching RMSD=5.8{\AA} at the end of the search. The average RMSD value over five runs at 400th iteration is 11.1$\pm$2.8{\AA}. We emphasise that better convergence could be reached if the MCTS is run longer. The best found conformer together with the target conformation are superimposed on each other, as is shown in Figure \ref{fig:5} (c). The best conformer and the target structure are quite similar. Exemplary transition between two distinct Fab arm conformations extracted by MCTS is shown in Figure \ref{fig:6}. The qualitative similarity with the one obtained by the non-targted run, is shown in Figure \ref{fig:1}.

\newpage

\section{Notes}

\begin{enumerate}
\item \emph{The input structure and its preprocessing}. Overall, the higher the resolution of the input structure, the more accurate motions will be produced by FRODAN. In order to extract accurate motions, the input structure should posses good covalent and steric geometry. Additional preprocessing before running FRODAN can potentially help to overcome this issue. Molecular-mechanics relaxation could be one of them. Unrealistic motions and geometry may also result from missing residues in the input structure. This can be fixed by using some residue adding software package.
\item  \emph{Absence of the force field}. FRODAN lacks a molecular mechanical force field. It employs constraint energy function instead. This is why FRODAN simulations may not be able to extract amount of details that conventional MD would. On the other hand, geometric sampling can rapidly produce stereochemically correct results. FRODAN can be very useful for large systems when using MD can becomes computationally demanding. Overall, it was shown that FRODAN generated motions are consistent with MD results \cite{farrell2011}.
\item  \emph{Absence of waters}.  Unfortunately FRODAN can not handle single-atom groups like waters, ions, etc.. This is why they must be removed from the structure before starting simulations.
\item  \emph{FRODAN conformational sampling limitations}. Wide conformational sampling using FRODAN non-targted mode is possible if the considered protein is highly flexible like antibody. However, when applied to more rigid structure, the conformational sampling will mostly be concentrated in the vicinity of initial state and distant conformations may be out of reach.
\item  \emph{MCTS set-up}. In order to use MCTS more systematically, non-trivial task is the selection of $\alpha$ and $C$ parameters. For the examples considered in this chapter, we used the default value $\alpha=1.1$ and $C=0.05$ from \cite{shin2019}. However, these values may not be optimal for achieving the best performance of the search when applied for the particular task.
\item  \emph{Search algorithm use perspectives}. In the example provided in this chapter, MCTS was used for transitioning between two distinct Fab-arm conformations, where we assume both initial and end structures are available. Nevertheless, MCTS as well as other search algorithms and their combinations can be applied to explore desired conformational space when 3D coordinates are unavailable, for example, fitting of cryo-electron maps with existing crystal structures.


\end{enumerate}

\bibliographystyle{unsrt}
\bibliography{mybib2}

\end{document}